\documentclass[fleqn,usenatbib]{mnras}

\usepackage{newtxtext,newtxmath}
\usepackage[T1]{fontenc}

\DeclareRobustCommand{\VAN}[3]{#2}
\let\VANthebibliography\thebibliography
\def\thebibliography{\DeclareRobustCommand{\VAN}[3]{##3}\VANthebibliography}

\usepackage{graphicx}	% Including figure files
\usepackage{amsmath}	% Advanced maths commands
\usepackage{CJK}
\usepackage{placeins}  % to bring figure A1 down below appendix a header
\newcommand{\vlsr}{\ensuremath{V_{\mathrm{LSR}}}}
\newcommand{\kms}{\ensuremath{\mathrm{\,km\,s^{-1}}}}
\newcommand{\kmspc}{\ensuremath{\mathrm{\,km\,s^{-1}\,pc^{-1}}}}
\newcommand{\Kkms}{\ensuremath{\mathrm{\,K\,km\,s^{-1}}}}
\title[FIRESTORM I: Feedback \& Kinematics in W40 HFS]
{FIRESTORM I: Stellar Feedback and Gas Kinematics in the Evolved W40 Hub-Filament System}
\author[Lim et al.]{Ming-Kang Lim$^{1}$\thanks{E-mail: mingkang@narit.or.th}, 
Ram K. Yadav$^{1}$\thanks{E-mail: ram$\_$kesh@narit.or.th},
L.~K. Dewangan$^{2}$,
Kee-Tae Kim$^{3,4}$,
A. Zavagno$^{5,6}$,
\newauthor
Jedsada Maklai$^{1,7}$,
Nicola Schneider$^{8}$,
D. Arzoumanian$^{9,10}$,
Arshia M. Jacob$^{8}$,
L.~E. Pirogov$^{11}$, 
\newauthor
Jihye Hwang$^{9,10}$, 
D.~K. Ojha$^{12}$,
Gyuho Lee$^{3,4}$,
Affan Adly Nazri$^{13}$, and 
Saurabh Sharma$^{14}$
\\
$^{1}$National Astronomical Research Institute of Thailand (Public Organization), 260 Moo 4, T. Donkaew, A. Maerim, Chiangmai 50180, Thailand.\\ 
$^{2}$Physical Research Laboratory, Navrangpura, Ahmedabad - 380 009, India.\\
$^{3}$Korea Astronomy and Space Science Institute, 776 Daedeokdae-ro, Yuseong-gu, Daejeon 34055, Republic of Korea.\\
$^{4}$University of Science and Technology, Korea (UST), 217 Gajeong-ro, Yuseong-gu, Daejeon 34113, Republic of Korea. \\
$^{5}$Aix-Marseille Universite, CNRS, CNES, LAM, Marseille, France.\\
$^{6}$Institut Universitaire de France, Paris, France.\\
$^{7}$Department of Physics and Materials Science, Faculty of Science, Chiang Mai University, Chiang Mai 50200, Thailand \\
$^{8}$I. Physikalisches Institut, Universität zu Köln, Zülpicher Str. 77, 50937 Köln, Germany.\\
$^{9}$The Institute for Advanced Study, Kyushu University \\
$^{10}$Department of Earth and Planetary Sciences, Faculty of Science, Kyushu University, Nishi-ku, Fukuoka 819-0395, Japan\\
$^{11}$Institute of Applied Physics of the Russian Academy of Sciences, 46 Ulyanov st., Nizhny Novgorod 603950, Russia.\\
$^{12}$Department of Astronomy and Astrophysics, Tata Institute of Fundamental Research, Homi Bhabha Road, Mumbai 400005, India.\\
$^{13}$Radio Cosmology Research Laboratory, Center for Astronomy and Astrophysics Research, Department of Physics, Faculty of Science,
Universiti Malaya, \\ 50603 Kuala Lumpur, Malaysia.\\
$^{14}$Aryabhatta Research Institute of Observational Sciences, Manora Peak, Nainital 263002, India.
}

\date{Accepted XXX. Received YYY; in original form ZZZ}

\pubyear{2025}

\begin{document}
\label{firstpage}
\pagerange{\pageref{firstpage}--\pageref{lastpage}}
\maketitle

\begin{abstract}
The FIRESTORM project--Feedback-Induced Regions and Emission from Star-forming Tracers of ObseRvable Molecular Gas--has targeted four star-forming regions to quantify the impact of stellar feedback on star formation. In this paper, we present multiwavelength results for one of the targets, the nearby high-mass star-forming region W40. Using dense-gas tracers C$^{18}$O(1--0) and H$^{13}$CO$^+$(1--0), we identified six velocity-coherent filaments: five at \vlsr $\sim$\,7.5\kms\! and one at \vlsr $\sim$\,5\kms. Four of these converge towards an infrared-bright cluster hosting the most massive star of the region (IRS 1A South, O9.5V), forming a hub-filament system (HFS). Key physical parameters, including filament lengths, widths, masses, velocity dispersions, and line masses, are derived. Five dense clumps traced by N$_2$H$^+$(1--0) exhibit subsonic to transonic turbulence, contrasting with the supersonic motions of their parental filaments, indicating turbulence dissipation. A deficit of emission at \vlsr $\sim$\,7\kms\! in several molecular lines, along with a blueshifted absorption dip in the HCN(1--0) profile, suggests that emission from OB-heated gas is being absorbed by a cold foreground cloud. A bridge-like feature in position-velocity space connects the \vlsr $\sim$\,5 and $\sim$\,7.5\kms\! filaments, and spatially coinciding with dense condensations and radio continuum peaks. These findings suggest that a past interaction--likely a cloud-cloud collision--triggered the formation of HFS and ultimately the central massive cluster. 
\end{abstract}


\section{Introduction} 
\label{sec:intro}
The formation of massive stars ($>$ 8 M$_{{\sun}}$) and stellar clusters remains one of the most fundamental yet poorly understood processes in astrophysics. Infrared and radio surveys have revealed that massive star-forming (SF) regions often harbour hub-filament systems \citep[HFSs;][]{myers2009, kirk2013, trevino-morales2019}, and there is a growing consensus that filamentary accretion within HFSs contributes significantly to mass assembly of massive stars and clusters \citep[e.g.,][]{motte2018, rosen2020, peretto2013}. In some configurations, massive stars may grow initially from low-mass stars formed at the hub \citep[e.g.,][]{motte2018}. HFSs are therefore key sites to understand the formation processes of massive stars and stellar groups \citep{Kumar_2020,suin2025}. 

Once massive stars form, their stellar feedback in the form of winds and radiation alters their environment. Extreme-UV radiation ionizes hydrogen and drives expanding \ion{H}{ii} regions \citep{stromgren1939, weaver1977, spitzer1978}, while far-UV radiation produces the surrounding photo-dissociation layers \citep{tielens1985,sternberg1989}. This feedback can reshape filaments and modify the initial conditions, promoting further star formation. 
Despite substantial progress, a number of unanswered questions remain, including: How do massive stars form? What leads to the formation of HFSs? And how do massive stars shape their surrounding interstellar medium (ISM) through feedback? 

\begin{figure*}
\center
\includegraphics[width=\textwidth]{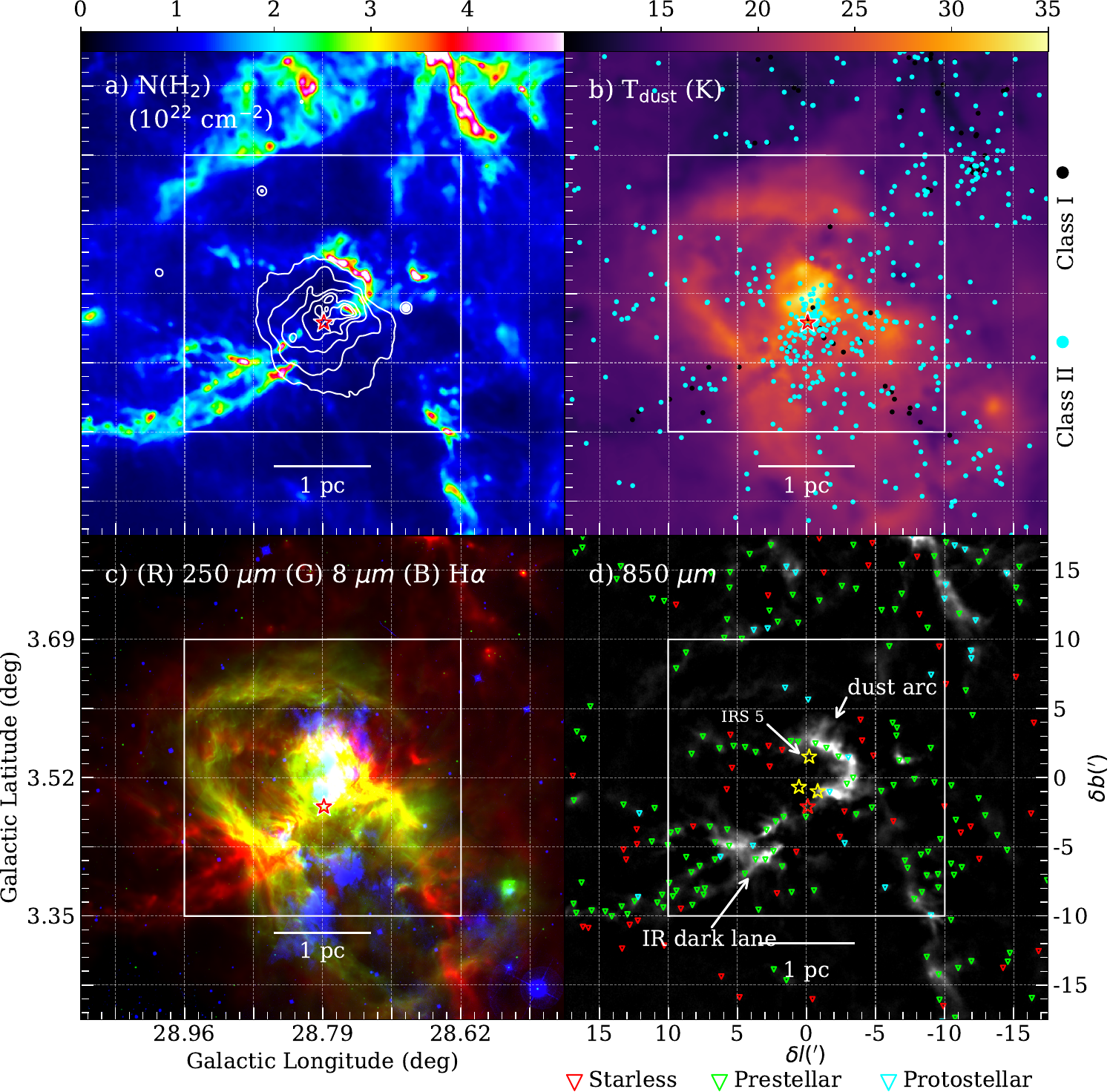}
\caption{(a) H$_2$ column density, $N_\mathrm{H_2}$ map from the 
    \emph{Herschel} Gould Belt Survey (HGBS) \citep{andre_2010} overlaid with 887.5 MHz contour from Rapid ASKAP Continuum Survey (RACS) Data Release 1. 
    Contour levels are [0.01, 0.05, 0.09, 0.13, 0.17, 0.21] Jy/beam. 
    (b) Dust temperature, $T_\text{dust}$ map from HGBS with overlaid positions of Class I (black) and II (cyan) YSOs from \citet{Sun_2022}. 
    (c) Composite image combining \emph{Herschel} 250\,\micron\, (red), \emph{Spitzer} 8\,\micron\, (green), and UKIRT H$\alpha$ emission (blue). 
    (d) JCMT SCUBA-2 850\,\micron\,, overlaid with \emph{Herschel} dense cores \citep{konyves2015}. 
    In each panel, the field of view of the TRAO observation is indicated by a white box, and the scale bar is drawn at a distance of 502 pc. Star symbols in panel (d) indicate the positions of massive stars identified by \citet{Shuping_2012}, with the most massive star, IRS 1A South, marked with a red star symbol in all the panels.}
\label{fig: w40_complex}
\end{figure*}

The SF complex W40 (see Fig.~\ref{fig: w40_complex}) is an ideal nearby laboratory for understanding these fundamental questions. At a distance of 502\,$\pm$\,4\,pc \citep{comeron2022}, adopted in this study, W40 harbours a bipolar \ion{H}{ii} region (Fig.~\ref{fig: w40_complex}b,c) with an angular diameter of $\sim$\,6\arcmin ($\sim$\,0.88 pc). The region is mainly ionized by the O9.5V star IRS 1A South \citep{Shuping_2012,comeron2022}. It has been recognized since early optical/radio surveys \citep{johnson1955, Westerhout_1958, sharpless1959}. Previous kinematic studies using H$\alpha$ \citep{crutcher1982} and \ion{C}{ii} \citep{faerber2025a} lines reveal expansion signature in the \ion{H}{ii} region. \emph{Herschel} imaging reveals a rich, multi-scale network of filaments (Fig.~\ref{fig: w40_complex}a) with a characteristic width of $\sim$\,0.1\,pc \citep{arzoumanian2011,arzoumanian2019}; the most supercritical filaments\footnote{Based on \citet{konyves2015}, a supercritical filament has its line mass (mass per unit length, $M_\mathrm{line}$) larger than its thermal critical line mass, $M_\mathrm{line,crit}$} are closely associated with dense cores \citep{andre_2010, menshchikov2010, konyves2015}. The clustering of young stellar objects (YSOs) (see Fig.~\ref{fig: w40_complex}b) further indicates multiple branches extending from the central cluster, with only a subset coincident with present-day molecular filaments, and suggests the partial disruption of the original HFS \citep{Sun_2022, sun2022}. On larger scales, W40 complex has been classified as an evolved (Stage IV) HFS exhibiting bubbles and pillars shaped by OB-star feedback \citep{Kumar_2020}. Its proximity and evolutionary state make it a prime site to test the interplay between filamentary inflow and stellar feedback. 

A velocity-resolved, multi-tracer filament analysis of the W40 cloud has not yet been carried out. But this analysis is the key to understand the dynamical interaction between the \ion{H}{ii} region and the surrounding SF region. We observed the W40 cloud as part of the FIRESTORM (Feedback-Induced Regions and Emission from Star-forming Tracers of ObseRvable Molecular Gas; PI: J. Maklai) program. FIRESTORM maps $J\! =\! 1\!-\!0$ 
rotational transitions of dense-gas tracers with the Taeduk Radio Astronomy Observatory (TRAO) 14-m across four representative Galactic SF regions--including W40--to investigate how OB-star feedback reshapes dense gas and regulates star formation. In this study, we combine dense-gas tracers\footnote{Unless otherwise stated, all molecular line transitions refer to the $J\! =\! 1\!-\!0$ transition.} and CO isotopologues--our TRAO 14-m observations of C$^{18}$O, N$_2$H$^+$, HCO$^+$, HCN, CN, and C$_2$H; together with Nobeyama 45-m $^{12}$CO, $^{13}$CO, and C$^{18}$O; IRAM 30-m H$^{13}$CO$^+$, and H$^{13}$CN; and JCMT $^{12}$CO(3--2) data--to connect structure and kinematics across the density hierarchy. Our analysis addresses the following key questions: (i) what physical processes assembled the W40 hub and filaments (converging flows, cloud-cloud interactions, or compression by feedback); and (ii) how feedback from the bipolar \ion{H}{ii}\ region has influenced the surrounding medium, including the possibility of triggering star formation in the region.

Section~\ref{sec: data} summarises observations and data reductions. 
Section~\ref{sec: analysis} presents the spatial and spectral distributions of the lines
alongside the identification of the filaments and clumps. 
Section~\ref{sec:discussion} discusses 
the role of feedback in star formation, and the origin of the W40 HFS. Finally, our summary and conclusions are given in Section~\ref{sec:conclusion}.

\section{OBSERVATIONS AND ANCILLARY DATA}
\label{sec: data}
A summary of the molecular line observations and archival data used in this paper is provided in Table~\ref{tab: molecular_line}. To approximate the density regime that each molecule traces, in the same table, we also listed the effective density, $n_\mathrm{eff}$ of the molecules, defined as the density of the collisional partner (typically H$_2$) which results in a molecular line with an integrated intensity of 1\,K\kms. We refer the reader to \citet{shirley2015} for an in-depth discussion of the use of $n_\mathrm{eff}$ over the more commonly used critical density, $n_\mathrm{crit}$. Several lines contain hyperfine structure splitting transitions that are resolvable by the spectrometer; their relative strengths and separations are listed in Table~\ref{tab: hfs_intensity_ratio}.

\begin{table*}
\centering
\caption{Summary of the molecular line observations used in this study}
\label{tab: molecular_line}
\begin{tabular}{lrlcccrccc}
\hline
\multicolumn{1}{c}{Molecular Line} &
\multicolumn{1}{c}{Frequency} & 
\multicolumn{1}{c}{Effective Density} &
\multicolumn{2}{c}{Beam Size} &
Channel Width &
\multicolumn{1}{c}{$T_\mathrm{peak}$} &
RMS Noise &
Facility &
Reference \\
\multicolumn{1}{c}{($J_u$--$J_l$)\textsuperscript{a}} &
\multicolumn{1}{c}{(GHz)} & 
\multicolumn{1}{c}{(log $n_\mathrm{eff}$) (cm$^{-3}$)} & 
($\theta_\mathrm{B}$) ($''$) &
($\theta_\mathrm{B}$) (pc) &
($\Delta v$) (\kms) &
\multicolumn{1}{c}{(K[$T_\mathrm{mb}$])}&
(K[$T_\mathrm{mb}$]) &
&
\\
\multicolumn{1}{c}{(1)} &
\multicolumn{1}{c}{(2)} &
\multicolumn{1}{c}{(3)} &
(4) &
(5) &
(6) &
\multicolumn{1}{c}{(7)} &
(8) &
(9) &
(10) \\
\hline
\multicolumn{10}{c}{\textbf{TRAO Observations}} \\
\hline
C$^{18}$O (1\,--\,0)                 & 109.782173 & 2.85\textsuperscript{c}            (14.99)\textsuperscript{d} & 49 & 0.12 & 0.042 & 6.2 & 0.66 & F1 & This work\\
CN        (1$_{3/2}$\,--\,0$_{1/2}$) & 113.490970 & 4.23\phantom{\textsuperscript{c}}  (14)\textsuperscript{b}    & 48 & 0.12 & 0.040 & 4.9 & 0.83 & F1 & This work\\
C$_2$H    (1$_{3/2}$\,--\,0$_{1/2}$) & 87.316925  & 5.29\textsuperscript{c}            (14)                       & 57 & 0.14 & 0.052 & 2.1 & 0.29 & F1 & This work\\
HCN       (1\,--\,0)                 & 88.631846  & 3.65\phantom{\textsuperscript{c}}  (14)\textsuperscript{b}    & 56 & 0.14 & 0.052 & 5.2 & 0.30 & F1 & This work\\
HCO$^+$   (1\,--\,0)                 & 89.188525  & 2.72\phantom{\textsuperscript{c}}  (14)\textsuperscript{b}    & 56 & 0.14 & 0.051 & 5.3 & 0.30 & F1 & This work\\
N$_2$H$^+$(1\,--\,0)                 & 93.173770  & 3.74\phantom{\textsuperscript{c}}  (13)\textsuperscript{b}    & 54 & 0.13 & 0.049 & 2.8 & 0.30 & F1 & This work\\
\hline 
\multicolumn{10}{c}{\textbf{Archival Data}} \\
\hline
% $^{12}$CO      (1\,--\,0)     & 115.271202    & ...                                                          & 21.7 & 0.05  & 0.100 & 52.60 & 0.70 & F2 & C1\\
% $^{13}$CO      (1\,--\,0)     & 110.201322    & 1.39\textsuperscript{c}           (16.33)\textsuperscript{d} & 22.1 & 0.05  & 0.100 & 35.04 & 0.26 & F2 & C1\\
% C$^{18}$O      (1\,--\,0)     & 109.782173    & 2.85\textsuperscript{c}           (14.99)\textsuperscript{d} & 22.2 & 0.05  & 0.100 &  8.99 & 0.36 & F2 & C1\\
% N$_2$H$^+$     (1\,--\,0)     & 93.173770     & 3.74\phantom{\textsuperscript{c}} (13)\textsuperscript{b}    & 24.1 & 0.06  & 0.100 &  6.97 & 0.31 & F2 & C1\\
% HCO$^+$        (1\,--\,0)     & 89.188525     & 2.72\phantom{\textsuperscript{c}} (14)\textsuperscript{b}    & 40.0 & 0.10  & 0.165 &  7.46 & 0.09 & F2 & C2\\
% H$^{13}$CO$^+$ (1\,--\,0)     & 86.754288     & 4.34\phantom{\textsuperscript{c}} (12.3)\textsuperscript{b}  & 40.0 & 0.10  & 0.165 &  1.74 & 0.07 & F3 & C2\\
% HCN            (1\,--\,0)     & 88.631846     & 3.65\phantom{\textsuperscript{c}} (14)\textsuperscript{b}    & 40.0 & 0.10  & 0.165 &  6.42 & 0.06 & F3 & C2\\
% H$^{13}$CN     (1\,--\,0)     & 86.340167     & 5.20\phantom{\textsuperscript{c}} (12.3)\textsuperscript{b}  & 40.0 & 0.10  & 0.165 &  0.67 & 0.06 & F3 & C2\\
% $^{12}$CO      (3\,--\,2)     & 345.795990    & ...                                                          & 14.6 & 0.04  & 0.050 & 37.50 & 1.36 & F4 & C3\\
$^{12}$CO      (1\,--\,0)     & 115.271202    & ...                                                          & 22 & 0.05  & 0.100 & 52.6 & 0.70 & F2 & C1\\
$^{13}$CO      (1\,--\,0)     & 110.201322    & 1.39\textsuperscript{c}           (16.33)\textsuperscript{d} & 22 & 0.05  & 0.100 & 35.0 & 0.26 & F2 & C1\\
C$^{18}$O      (1\,--\,0)     & 109.782173    & 2.85\textsuperscript{c}           (14.99)\textsuperscript{d} & 22 & 0.05  & 0.100 & 9.0 & 0.36 & F2 & C1\\
N$_2$H$^+$     (1\,--\,0)     & 93.173770     & 3.74\phantom{\textsuperscript{c}} (13)\textsuperscript{b}    & 24 & 0.06  & 0.100 & 7.0 & 0.31 & F2 & C1\\
HCO$^+$        (1\,--\,0)     & 89.188525     & 2.72\phantom{\textsuperscript{c}} (14)\textsuperscript{b}    & 40 & 0.10  & 0.165 & 7.5 & 0.09 & F3 & C2\\
H$^{13}$CO$^+$ (1\,--\,0)     & 86.754288     & 4.34\phantom{\textsuperscript{c}} (12.3)\textsuperscript{b}  & 40 & 0.10  & 0.165 & 1.7 & 0.07 & F3 & C2\\
HCN            (1\,--\,0)     & 88.631846     & 3.65\phantom{\textsuperscript{c}} (14)\textsuperscript{b}    & 40 & 0.10  & 0.165 & 6.4 & 0.06 & F3 & C2\\
H$^{13}$CN     (1\,--\,0)     & 86.340167     & 5.20\phantom{\textsuperscript{c}} (12.3)\textsuperscript{b}  & 40 & 0.10  & 0.165 & 0.7 & 0.06 & F3 & C2\\
$^{12}$CO      (3\,--\,2)     & 345.795990    & ...                                                          & 15 & 0.04  & 0.050 & 80.7 & 1.90 & F4 & C3\\
\hline
\multicolumn{10}{l}{\textsuperscript{a} For species with fine structure (CN and C$_2$H), the notation is ($N_{uJ_u}$--$N_{lJ_l}$)}\\
\multicolumn{10}{l}{\textsuperscript{b} Values of $n_\mathrm{eff}$ and $N_\mathrm{ref}$ are adopted from Table 1 in \cite{shirley2015}}\\
\multicolumn{10}{l}{\textsuperscript{c} The value of $n_\mathrm{eff}$ is calculated using \texttt{spectralradex} radiative transfer code \citep{holdship2021}}\\
\multicolumn{10}{l}{\textsuperscript{d} The $N_\mathrm{ref}$ values of $^{13}$CO(1--0) and C$^{18}$O(1--0) correspond to the mean $N_\mathrm{ref}$ values in the W40 cloud %value
reported in \cite{tursun2024}}\\
\hline
\end{tabular}
\medskip
\begin{minipage}{0.97\textwidth}
\textit{Column descriptions}: (1) Quantum numbers of the transition based on the notation provided in the Cologne Database for Molecular Spectroscopy (CDMS; \citealt{muller2005}); (2) Frequency of the transition; (3) Effective density of the transition assuming a kinetic temperature $T_k$ of 20\,K. The value in parenthesis refers to the reference column density in log scale (log $N_\mathrm{ref}$) of which $n_\mathrm{eff}$ is calculated; (4) Angular resolution of the final data cube in arcsec; (5) Angular resolution of the final data cube in pc at the distance of 502\,pc; (6) Channel width of the final data cube (7) Peak main beam temperature within the observed area; (8) Mean rms noise measured in a emission-free channels (e.g., first 100 channels); (9) Facilities: F1 = TRAO14-m SEQUOIA, F2 = NRO45-m FOREST, F3 = IRAM30-m EMIR, F4 = JCMT15-m HARP; (10) References: C1 = \citet{nakamura2019}, C2 = \citet{shimajiri_2017}, C3 = \citet{rumble2016}.
\end{minipage}
\end{table*}

\subsection{Observations with the TRAO 14-m Telescope} \label{subsec:observation}
As part of the large program FIRESTORM (PI: Maklai, J.), we mapped the W40 cloud in the $J\! =\! 1\!-\!0$ rotational transitions of C$^{18}$O, N$_2$H$^+$, HCO$^+$, HCN, CN, and C$_2$H lines between 87 and 114 GHz using the TRAO 14-m telescope with the Second Quabbin Optical Image Array (SEQUOIA), a 4×4 MMIC preamplifier receiver array system \citep{Jeong_2019}. The observations were carried out in the on-the-fly (OTF) mode from late 2022 to early 2023, using an FFT2G spectrometer as the backend. For individual spectra, the total bandwidth and the number of channels were respectively 62.5\,MHz (corresponding to 187\kms\! at 100\,GHz) and 4096, resulting in a spectral resolution of 15\,kHz (0.045\kms\! velocity resolution at 100\,GHz). Full observational details will be provided in a forthcoming paper (Yadav et al., in prep). The mapped area covers 20\arcmin\,$\times$\,20\arcmin (2.92\,pc\,$\times$\,2.92\,pc), centred at ($l$\,=\,28.79\degr, $b$\,=\,3.52\degr), chosen to encompass the HFS in the W40 complex. The pointing and focus of the telescope were checked out every $\sim$\,3 hours by observing strong SiO maser sources. The pointing accuracy was better than 10\arcsec. The OFF position is ($l$\,=\,28.79\degr, $b$\,=\,4.52\degr), which does not show any appreciable HCO$^+$ and HCN line emission. The data were calibrated by the standard chopper wheel method. The line intensity was initially measured on the antenna temperature ($T^*_\text{A}$) scale and later converted to the main beam brightness temperature, $T_{\text{mb}}$ using the equation $T_{\text{mb}}=T^*_\text{A}/\eta_\text{B}$, where $\eta_\text{B}$ denotes the telescope's main beam efficiency. The main beam size ranges from  46\arcsec to 57\arcsec, and the corresponding $\eta_\text{B}$ values range from 0.42 to 0.50, as reported in the \href{https://trao.kasi.re.kr/status_report.php?id=2020}{TRAO 2020 status report}. All the OTF scans were baseline corrected using second-order polynomial fitting, regridded to a pixel size of 20\arcsec\,$\times$\,20\arcsec using \textsc{otftools}, and subsequently made into \texttt{FITS} cubes using \textsc{gildas class}. The average rms noises in $T_\text{mb}$ across the final data cubes range from 0.29\,--\,0.83\,K 
among all observed lines. The mean system temperatures, $T_\mathrm{sys}$ are $\sim$\,178\,K for N$_2$H$^+$, $\sim$\,185\,K for HCN, HCO$^+$ and C$_2$H, 251\,K for C$^{18}$O and 375\,K for CN. 

\subsection{Ancillary Data} \label{subsec:ancillary_data}
In addition to the TRAO observations, we incorporated archival molecular line data sets, continuum images, and source catalogues to complement our analysis. All data sets were cropped to a 40\arcmin\,$\times$\,40\arcmin\! region centred at ($l$\,=\,28.79\degr, $b$\,=\,3.52\degr), which encompasses both the filamentary dust structure and the associated bipolar infrared morphology.

\subsubsection{Molecular Line Spectroscopy}
The W40 and the nearby young SF region Serpens South were mapped in the $J\! =\! 1\!-\!0$ transition of $^{12}$CO, $^{13}$CO, C$^{18}$O, N$_2$H$^+$ species and CCS(J$_N$=$8_7$--$7_6$) using the Nobeyama 45-m telescope, covering a field of view (FoV) of 1\,deg$^2$ and a beam size, $\theta_\mathrm{B}$ of 22\farcs2 (0.054\,pc) at 109\,GHz \citep{Nakamura_2019}. The same data were analysed by \citet{shimoikura2019,shimoikura2020}. We retrieved all data except CCS(J$_N$=$8_7$--$7_6$)\footnote{CCS(J$_N$=$8_7$--$7_6$) is undetected in the W40 region} in $T_\mathrm{mb}$ unit from the \href{https://jvo.nao.ac.jp/portal/nobeyama/sfp.do}{JVO portal} operated by ADC/NAOJ, based on observations from the Nobeyama Radio Observatory (NRO). 

The W40 and Serpens South regions were also mapped in the \emph{J}=1--0 transition of HCO$^+$, H$^{13}$CO$^+$, HCN, and H$^{13}$CN by IRAM 30-m telescope, with a beam size, $\theta_\mathrm{B}$ of 28\farcs6 (0.07\,pc) at 86\,GHz \citep{shimajiri_2017}. These data cubes, provided in the $T_\mathrm{mb}$ unit, were smoothed to an angular resolution of 40\arcsec\! and obtained via private communication with Y. Shimajiri.

The W40 region was mapped in the $^{12}$CO(3--2) using the HARP receiver on JCMT \citep{rumble2016} (Proposal ID: M15AI31), with a beam size $\theta_\mathrm{B}$ of 14\farcs6 and a FoV of 7\arcmin\,$\times$\,18\arcmin, confined to the area covering the western dust arc and partially the eastern infrared dark lane. The observations are available in two different channel spacings: 0.061 and 0.977 MHz (0.05 and 0.85\kms\! at 346 GHz, respectively). We retrieved the data with the finer channel spacing, provided in the $T_A^*$ unit, and converted to $T_\mathrm{mb}$ by dividing the data by 0.61, the main beam efficiency reported in \citet{rumble2016}.

We cross-checked the TRAO C$^{18}$O and N$_2$H$^+$ data with those from NRO, and HCN and HCO$^+$ data with those from IRAM. Before comparison, all data sets were smoothed using Gaussian kernels to achieve a common spatial and spectral resolutions. The kernel widths were calculated as $\sqrt{\theta_\mathrm{target}^2-\theta_\mathrm{cube}^2}$ for spatial smoothing and $\sqrt{\Delta v_\mathrm{target}^2-\Delta v_\mathrm{cube}^2}$ for spectral smoothing. 
We found that the TRAO C$^{18}$O data are systematically weaker ($\sim$1\,K or 39\%) than the NRO at $\sim$\,7\kms\!. At some positions TRAO shows no emission or apparent absorption, while NRO shows emission of $\sim$1\,K\,($T_\mathrm{mb}$). In other positions with a 7\kms\! peak, both telescopes detect emission, but NRO is brighter by $\sim$1--1.5\,K (see Fig.~\ref{fig: c18o_nro_trao}). To test whether this discrepancy arises from OFF-position contamination, we inspected Purple Mountain Observatory (PMO) archival OTF maps that cover the sky position used as TRAO's OFF reference. Those PMO maps show average $^{12}$CO and $^{13}$CO emission of $T^*_A\approx2\,\mathrm{K}$ at $\vlsr\approx7.5\kms$, and one map shows weak C$^{18}$O emission ($T^*_A\approx0.4\,\mathrm{K}$) at the same velocity. Contamination in the OFF position may therefore contribute to the inconsistency in the spectra.

The N$_2$H$^+$, HCO$^+$, and HCN data agree with the archival data sets within a mean rms difference of 0.24\,K (8.36\,\%), 0.42\,K (6.75\,\%), and 0.18\,K (4.39\,\%), respectively. 
The $T_\mathrm{peak}$ ratios ($T\mathrm{_{peak}^{archival}}/T\mathrm{_{peak}^{TRAO}}$) range from 0.5 to 1.5 for all three lines, with median values of $\sim$0.95. Approximately 68\% of data points lie between ratios 0.8 and 1.1, indicating overall good agreement with archival data. Elevated ratios are found towards some clumpy regions identified in previous millimetre studies \citep[e.g.,][]{maury2011,Pirogov_2013}. These differences are likely attributable to calibration uncertainties between telescopes (e.g., conversion from $T^*_A$ to $T_\mathrm{mb}$), beam coupling effects due to compact structures, and additional errors introduced by Gaussian smoothing and baseline fitting.

\subsubsection{Continuum and others}
We obtained the 8\,\micron\ image ($\theta_\mathrm{B}=1\farcs71$ or $4.16\times10^{-3}$\,pc) 
from the {\it Spitzer} Enhanced Imaging Products (SEIP) \citep{seip}, 
and H$\alpha$ image ($\theta_\mathrm{B}$\,$\sim$\,1$^{\prime\prime}$ or $2.43\times 10^{-3}\,\mathrm{pc}$) at 6548\,\text{\AA} from the \href{http://www-wfau.roe.ac.uk/sss/halpha/hapixel.html}{UKST SuperCOSMOS $\mathrm{H}\alpha$ survey} \citep{parker_2005}. The \emph{Herschel} column density ($N_\mathrm{H_2}$) and dust temperature ($T_\mathrm{dust}$) maps ($\theta_\text{B}$=18\farcs2 or 0.044\,pc), together with the 250\,\micron\, image and the dense cores catalogue, were obtained from \href{http://www.herschel.fr/cea/gouldbelt/en/Phocea/Vie_des_labos/Ast/ast_visu.php?id_ast=66}{\emph{Herschel} Gould Belt Survey (HGBS)} \citep{andre_2010, konyves2015}. The 887.5 MHz ($\theta_\mathrm{B}=15\arcsec$ or 0.037\,pc) radio continuum image was obtained from the Rapid ASKAP Continuum Survey (RACS) Data Release 1 \citep{mcconnell2020}. The 850\,\micron\, ($\theta_\mathrm{B}=14\farcs8$ or 0.036\,pc) image was retrieved from the JCMT Gould Belt Survey \citep{ward-thompson2007a,kirk2018}.

\section{Results}
\label{sec: analysis}
Fig.~\ref{fig: w40_complex} shows the W40 complex  
across multiple wavelengths. For consistency, we adopt an offset coordinate system centred at ($l$\,=\,28.79\degr, $b$\,=\,3.52\degr), which will be used throughout the paper. In panel (a), the main ionizing source, the O9.5V star IRS 1A South, lies near the centre of the 887.5 MHz radio continuum, which is approximately symmetric about the waist (oriented northwest-southeast) of the bipolar morphology. Other ionizing sources include a B1 and two B2 stars \citep{comeron2022}. The radio continuum contours exhibit a noticeable pinched morphology towards the northwest and southeast edges of the \ion{H}{ii} region. These distortions coincide spatially with dense filamentary structures seen in the dust emission, suggesting that the expansion of the ionized bubble is impeded in these directions, where the surrounding gas density is higher. This interaction leads to an asymmetric shell: the \ion{H}{ii} region preferentially expands into regions of lower ambient density, while being confined or slowed down where it encounters the filaments. 

In panel (b), the dust temperature map also exhibits a bipolar morphology. Class II YSOs are clustered around IRS 1A South, whereas Class I YSOs are predominantly distributed along the southeastern filamentary structures and southwestern  
clumpy dust structures within the southern lobe of the bipolar \ion{H}{ii}\ region. 

In panel (c), H$\alpha$ emission appears patchy and extends primarily in the north–south direction. The 8\,\micron\, emission traces a bipolar structure, while 250\,\micron\, emission highlights prominent filamentary features to the southeastern and southwestern directions.

In panel (d), the 850\,\micron\, continuum traces cold and dense molecular clouds similar to the highest column density regions in the \emph{Herschel} column density map. Extended emission is filtered out by the JCMT data reduction process \citep{rumble2016}, leaving mainly high density structures concentrated along the boundary of the radio continuum. Their distribution suggests density enhancements caused by compression from the expanding \ion{H}{ii}\ region.

\subsection{Molecular Cloud Distribution and Self-Absorption} 
\label{subsec: molecular_cloud_distribution}
\subsubsection{Molecular Cloud Distribution} \label{subsubsec: molecular_cloud_distribution}
The W40 cloud has been reported to 
exhibit complex velocity structures, with multiple  
components 
peaking at approximately 3, 5, 7, 8, and 10\kms\! \citep[see Figure~18 in][]{shimoikura2019}. Based on the averaged spectra across the region in several lines (see Appendix
~\ref{sec: avg_spectra}), we detected four main velocity components (3, 5, 7, and 10\kms). A $\sim$\,40\kms\! high-velocity feature has also been reported in \cite{shimoikura2020}, but we do not analyze it here because it shows no kinematic connection with the other components in position-velocity space (Figure~8 of \citealt{shimoikura2020}). We therefore consider it a foreground or background cloud along the line of sight.

\begin{figure*} 
    \centering
    \includegraphics[width=0.88\textwidth]{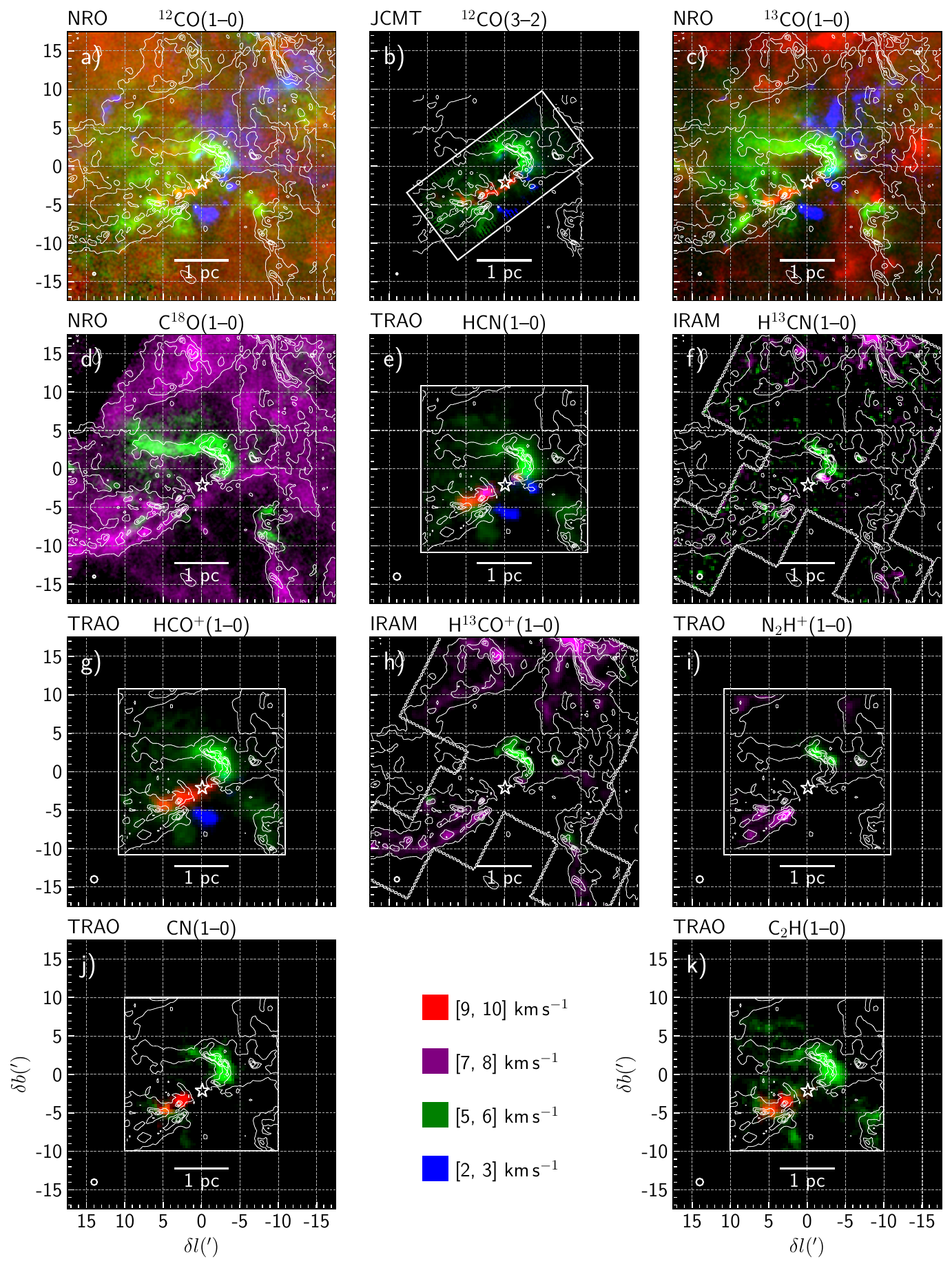}
    \caption{Colour composite moment-0 maps of molecular lines. Each map shares the same FoV and spatial axes. Each colour represents a moment-0 map integrated over the velocity ranges indicated in the figure\textsuperscript{a}. These ranges were selected to represent the four main velocity components associated with the W40 cloud
    (3, 5, 7, and 10\kms). 
    The white boxes in panels b), e), f), g), h), i), j), and k) indicate the extent of the corresponding map. The FoV of each map is the same as in Fig.~\ref{fig: w40_complex}. White circle in the lower left corner of each panel denotes the angular resolution of the respective data cube (see Table~\ref{tab: molecular_line} for the resolutions). The location of IRS 1A South is marked with a star symbol in each panel. White contours show the \emph{Herschel} column density map with contour levels of [1, 2, 3, 4] $\times10^{22}\,\mathrm{cm}^{-2}$. The facility of which the data is obtained is indicated at the upper left corner of the respective panel.}
    \begin{minipage}{1\textwidth}
    \raggedright
    \footnotesize
    \textsuperscript{a} Except HCN, where the range [13, 14]\kms\! is used for red channel due to overlap between F(1--1) hyperfine line of the 5\kms\! component 
    and the F(2--1) line of 10\kms\! component 
    within the [9, 10]\kms\! interval.
    \end{minipage}
    \label{fig: w40_moment0_maps_rgb}
\end{figure*}

Fig.~\ref{fig: w40_moment0_maps_rgb} presents the 
RGB composite maps of integrated intensity, generated by intensity integration over selected 1\kms\! velocity bins. These maps reveal the presence of multiple velocity components in each molecular line. The velocity bins are selected based on visually identified peaks in spectra averaged across the observed area (see Appendix~\ref{sec: avg_spectra}). An integration width of 1\kms\! was chosen as a compromise between achieving sufficient signal for map construction and minimizing contamination from nearby velocity components. For lines with resolvable hyperfine structure, only the main (strongest) line is considered.

All moment maps in this study were generated using the \texttt{Behind the Spectrum (BTS)} code\footnote{Except for H$^{13}$CN, where the emission is too weak ($\sim$\,0.5\,K\kms), so the moment-0 map was computed without masking.} developed by \citet{Clarke_2018}, which produces cleaner maps--particularly for data with weak emission (e.g., CN and C$_2$H). This code implements the moment-masking technique, following the method described by \cite{Dame_2011}, to effectively reduce noise while preserving weak emission signals. The algorithm first unmasks pixels whose emission exceeds a temperature threshold $T_C$. It then iteratively unmasks neighbouring pixels with emission above a lower threshold $T_L$, continuing this process until no additional pixels satisfy this criterion. We set the default thresholds as $T_C = 8\sigma$ and $T_L = 5\sigma$, where $\sigma$ is the rms of the emission-free channels. 

Comparing the line tracers in Fig.~\ref{fig: w40_moment0_maps_rgb}
, differences in the emission distributions because of temperature effects, density variations, and chemistry become obvious. The full, mostly filamentary, molecular cloud structure is best seen in C$^{18}$O, which resembles most the \emph{Herschel} column density map. The $^{12}$CO and $^{13}$CO maps miss some emission features that are visible in the \emph{Herschel} column density map (contours in Fig.~\ref{fig: w40_moment0_maps_rgb}) and in the C$^{18}$O map (Fig.~\ref{fig: w40_moment0_maps_rgb}d). It is mainly the 7\kms\! component, corresponding to the bulk emission of the cloud, is clearly seen in C$^{18}$O emission and extends farther outwards from IRS 1A South, but is missing in the $^{12}$CO and $^{13}$CO lines. We will show in Section~\ref{subsubsec: self-absorption} that this is mostly due to self-absorption of this velocity component in the $^{12}$CO--and partly in the $^{13}$CO--lines
. The optically thin lines of H$^{13}$CO$^+$, H$^{13}$CN also trace the main filamentary structures
in the W40 cloud, but are less sensitive to the lower-density gas. N$_2$H$^+$ is another high-density tracer, but it additionally focuses on cold gas (typically $<$10--20\,K). Interestingly, it outlines the more clumpy--and not filamentary--structure of the molecular gas. In contrast, the warmer molecular gas and the photodissociation region (PDR) close to the central exciting star are best seen in the important chemistry tracers CN, HCN, HCO$^+$, and C$_2$H. Most prominent is a bright clump northwest of IRS\,1A\,South. An additional bright emission peak southeast of the central source is only seen in C$_2$H. C$_2$H traces best UV-irradiated, moderately dense molecular gas. CN is an equally good PDR tracer, but requires higher densities. The southeastern clump, thus, is probably less dense than the northwestern one. In the following, we discuss in more detail the different velocity components found in the W40 cloud
.

The velocity structures differ significantly between tracers. For example, the 7\kms\! component 
is primarily detected in C$^{18}$O, H$^{13}$CN, H$^{13}$CO$^+$, and N$_2$H$^+$. These tracers show no evidence of the 10\kms\! component, which instead appears only in the other lines. On the other hand, the 3\kms\! component 
is only detected in the $^{12}$CO, $^{12}$CO(3--2), $^{13}$CO, HCN and HCO$^+$.

In terms of spatial distribution, the 7\kms\! component 
is located mainly outside the edge of the bipolar \ion{H}{ii} bubble, and shows a filamentary \footnote{Note that this is only a qualitative description, we performed a dedicated filament detection in Section~\ref{sec: filament_analysis}} structure extending towards the southeast. The 5\kms\! component 
lies northwest of the massive stars and extends largely within the bubble. The 10\kms\! component 
is concentrated at the bubble's waist, while the 3\kms\! component 
appears clumpy and is distributed south of the massive stars, at the southern lobe of the bipolar bubble.

\subsubsection{Self-Absorption} \label{subsubsec: self-absorption}
We suspect that the non-detection of the 7 km s$^{-1}$ component 
in most tracers is due to self-absorption. This occurs in the presence of two layers within the 7\kms\! component 
with different temperatures (or a temperature gradient), where a cold, dense foreground layer 
absorbs the emission from background warmer gas, resulting in their suppression or non-detection. This phenomenon is reported in previous studies in the W40 cloud, based on comparisons between optically thick lines and their optically thin isotopologues \citep{Pirogov_2013, shimoikura2015, shimajiri_2017, shimoikura2019, komesh2020}. In particular, the 7\kms\! component appears weak in $^{12}$CO and is absent in HCO$^+$, while it is clearly detected as a peak in C$^{18}$O and H$^{13}$CO$^+$. Similar self-absorption features have also been observed in the nearby Serpens South region \citep{nakamura2011}.

HCN with its hyperfine structure, offers a new opportunity to investigate this effect, since the degree of self-absorption depends on the line opacity, which differs among hyperfine lines. Additionally, for the 2\kms\! component
, the satellite HCN($J\!=\!1\!-\!0, F\!=\!1\!-\!1$) line falls at $\sim$7\kms, so we can see its line behaviour. 

\begin{figure*}
    \centering
    \includegraphics[width=\textwidth]{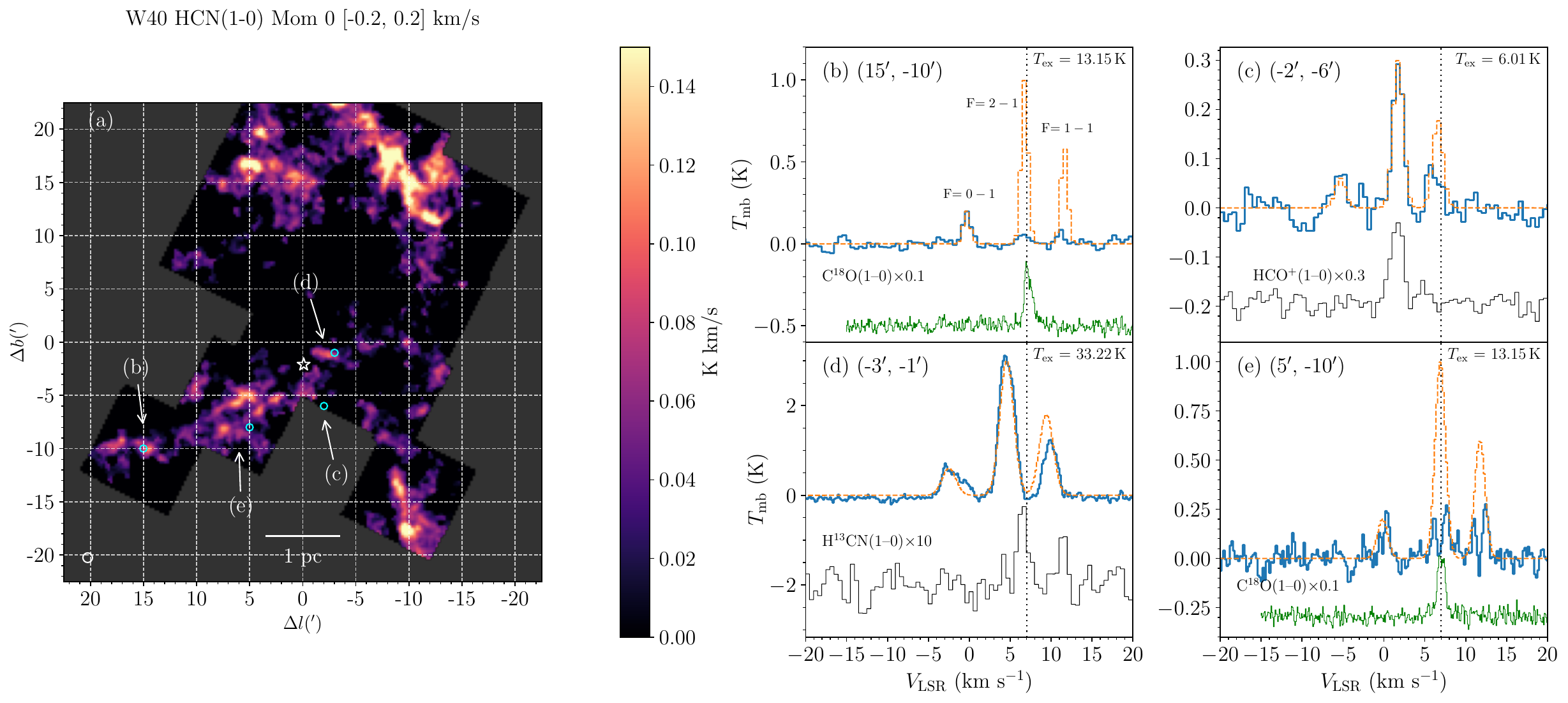}
    \caption{(a) Integrated intensity map of HCN over the velocity range [--0.2, +0.2]\kms, centred on the F(0--1) hyperfine line of the 7\kms\! component. Filamentary structures in the southeastern region (labelled SE1 and SE2 in Fig.~\ref{fig: filament_vel_info}a) are clearly visible. (b--e) HCN spectra at selected positions, illustrating the effect of foreground absorption at 7\kms, with spectra of C$^{18}$O, HCO$^+$ and H$^{13}$CN at the same positions as references. The orange dotted spectrum in each panel represents the expected profile under Local Thermodynamic Equilibrium (LTE) and optically thin ($\tau_m=0.1$) conditions, with excitation temperature $T_\mathrm{ex}$ value indicated at the upper right of each panel. The transitions of HCN hyperfine lines are indicated in panel (b). (b) At the 7\kms\! component
    , the F(2--1) and F(1--1) lines are completely absorbed, while the F(0--1) line is less affected. (c) At the 3\kms\! component
    , the F(1--1) line (which lies at 7\kms) shows suppressed emission due to absorption. (d) At the 5\kms\! component
    , the F(2--1) and F(1--1) lines exhibit steeper red wings, suggesting partial absorption. (e) At the 7\kms\! component
    , a blueshifted dip is observed, indicating possible expansion of the absorbing foreground layer. The HCN spectra in panels (b) and (c), as well as the H$^{13}$CN spectrum, are smoothed to 0.5\kms, while the spectra in panel (e) are smoothed to 0.2\kms.} 
    \label{fig: hcn_absorption_spectra}
\end{figure*}

Careful inspection of the HCN spectra reveals several noteworthy features. A few examples of spectra with distinct profiles are shown in Fig.~\ref{fig: hcn_absorption_spectra}, and are described below:
\begin{enumerate}
    \item In the 7\kms\! component
    , at specific positions (e.g., offset by ($+5',-10'$) from the central position defined in Section~\ref{sec: analysis}), the HCN spectra are not fully absorbed and instead show a blueshifted dip (See Fig.~\ref{fig: hcn_absorption_spectra}d), suggesting that the absorbing foreground cloud is expanding towards the observer. Interestingly, the HCN(J=1--0, F=0--1) hyperfine line is less affected by self-absorption (see Fig.~\ref{fig: hcn_absorption_spectra}a), likely due to its lower optical depth (20\% of the main F(2--1) line). As a result, filamentary structures are visible in the F(0--1) line. Fig.~\ref{fig: hcn_absorption_spectra}a) shows the moment-0 map of F(0--1) line, where filamentary structures are clearly seen, but are absent in F(1--1) and F(2--1) lines (not shown).
    \item In the 5\kms\! component
    , the F(1--1) line appears suppressed (see Fig.~\ref{fig: hcn_absorption_spectra}b and Fig.~\ref{fig: hcn_hfs_ratio}). This behaviour can be attributed to a combination of two effects: (1) hyperfine anomalies, in which the relative intensity of the F(1--1) line comparing with the main F(2--1) line is lower than the LTE range (0.6--1) due to population transfer from line overlap effect in the higher rotational transitions \citep{Goicoechea_2022}, and (2) self-absorption, the F(2--1) line often shows steeper red wing,\footnote{The mean rms of the residuals after Gaussian fitting at velocity interval of [6--8]\kms\! (red wing) is higher than that in [2--4]\kms\! (blue wing).} which can be due to self-absorption by the 7\kms\! component that partially attenuates the F(2--1) line.
    \item In the 3\kms\! component
    , the F(1--1) hyperfine line (near 7\kms) is absent (see Fig.~\ref{fig: hcn_absorption_spectra}b).
\end{enumerate}

Taken together, the HCN spectra provide additional evidence for self-absorption in the 7\kms\! component 
in the W40 cloud. The fact that this component is ubiquitous in C$^{18}$O implies that it is relatively dense, to the extent that emission from some species is entirely absorbed.

\subsection{Filament Identification and Analyses} \label{sec: filament_analysis}
Since C$^{18}$O and H$^{13}$CO$^+$ are not significantly (or only weakly) affected by self-absorption, and their integrated intensity maps reveal filamentary structures similar to those seen in dust emission, we can use their velocity information to identify filaments as velocity-coherent structures. Filaments
have been successfully detected in these transitions in other regions (C$^{18}$O: \citealt{orkisz2019, suri2019, trevino-morales2019, liu2021a}; H$^{13}$CO$^+$: \citealt{schneider2010, yang2023}). This allows a direct comparison of filament properties derived from different tracers.

We used \textsc{astrodendro} \citep{rosolowsky2008} to identify 
 filaments in NRO C$^{18}$O and IRAM H$^{13}$CO$^+$ data. This tool is widely employed to characterise hierarchical structures such as filaments \citep[e.g.,][]{shimajiri2023, Chung_2019}, clumps \citep[e.g.,][]{watkins2019}, and cores \citep[e.g.,][]{omodaka2020} using a dendrogram-based algorithm. \citet{chira2018} compared several filament-finding algorithms applied to simulations and reported large differences in the detailed structures they recover, though they agree well on the brightest filaments—the focus of our analysis. The identification procedures are detailed in Appendix~\ref{sec: filament_identification}.

In total, we identified three filaments (labelled as N1, W1 and W2) from the C$^{18}$O data, and three filaments (labelled as N2, SE1 and SE2) from the H$^{13}$CO$^+$ data. SE1 is also detected in C$^{18}$O. However, H$^{13}$CO$^+$ is chosen to perform the analysis due to its simpler spectral profile. Nonetheless, this provides an opportunity to assess the differences in filament properties derived using different tracers (e.g., width, see Section~\ref{sec: filament_width}).  

\begin{figure*} 
    \centering
    \includegraphics[width=\textwidth]{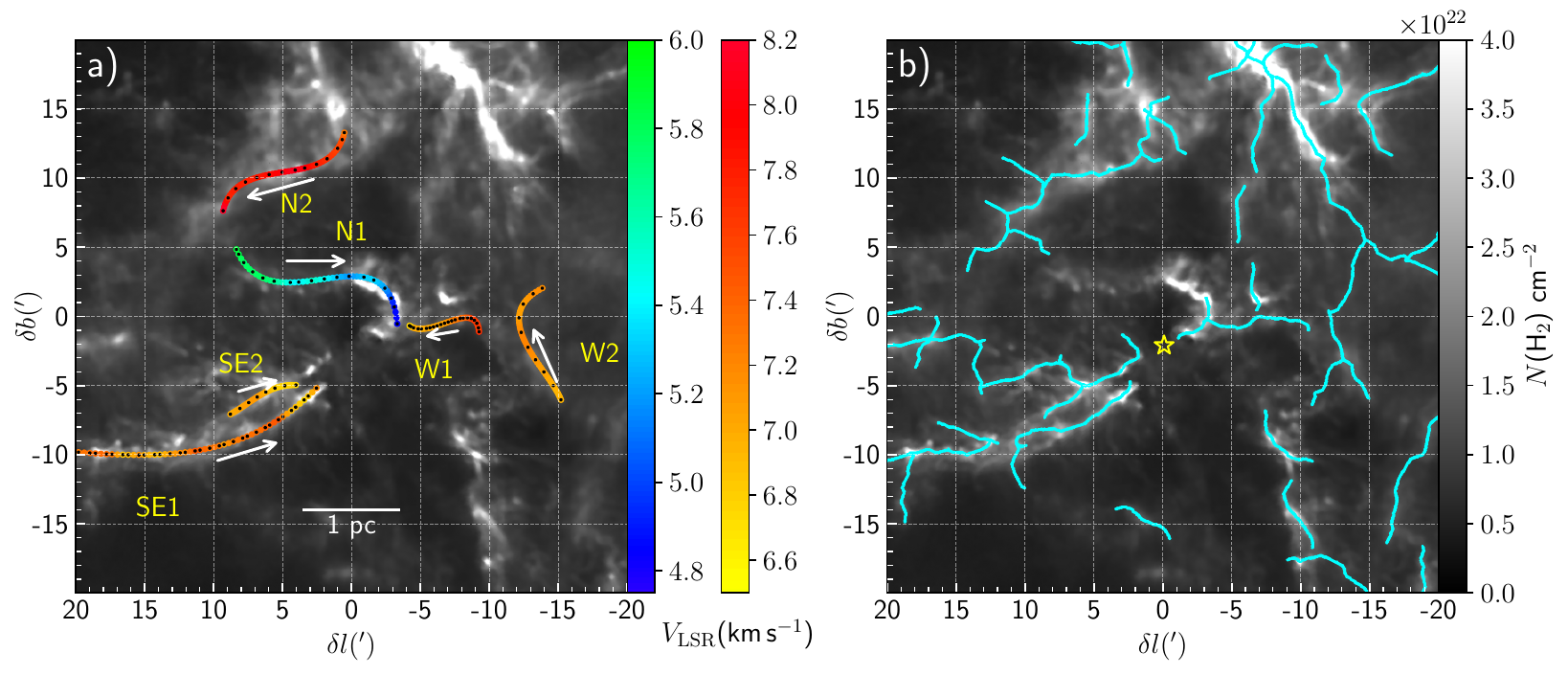}
    \caption{(a) 
    Spatial distribution of filament skeletons analysed in this study. The skeletons are colour-coded with their $\vlsr$ profiles, linearly interpolated from the sampled points (black dots). Each filament is labelled with its designation. White arrows illustrate the direction of increasing filament length (corresponding to the positive x-axis in Fig.~\ref{fig: filament_T_vlsr_dv_profile}). Note that N1 is shown with a distinct colourmap for clarity. (b) Skeletons of filaments detected from HGBS \citep{konyves2015}. Both panels share the same FoV and background image (\emph{Herschel} column density map). The scale bar and star symbol in each panel follow the definitions in Fig.~\ref{fig: w40_complex}.}
    \label{fig: filament_vel_info}
\end{figure*}

We skeletonized the 2D filament mask using \texttt{skimage.morphology.skeletonize}, and retained the longest path within each connected component to define the filament skeleton. The spatial distribution of the filament skeletons is shown in Fig.~\ref{fig: filament_vel_info}. In the same figure, we also plotted the filament skeletons identified in the HGBS survey \citep{konyves2015}. A comparison shows that SE1, identified here as a single structure, was separated into two distinct filaments in HGBS, while N1 was not reported in that survey. N2 and W1 appear longer in HGBS, and W1 in particular seems to extend towards Serpens South. In contrast, several filaments in the eastern and southern regions were identified in HGBS but are not detected in our analysis.

We derived filament lengths and widths without correcting for projection effects, using the publicly available \texttt{FilChaP} package \citep{suri2019} (see Appendix~\ref{sec: filament_length_and_width_derived}). The values are derived from the same tracer in which the filament is identified. True lengths are expected to be larger by factor of $\approx$1.2--1.6 on average, assuming random orientations.

\begin{figure*}
    \centering
    \includegraphics[width=\textwidth]{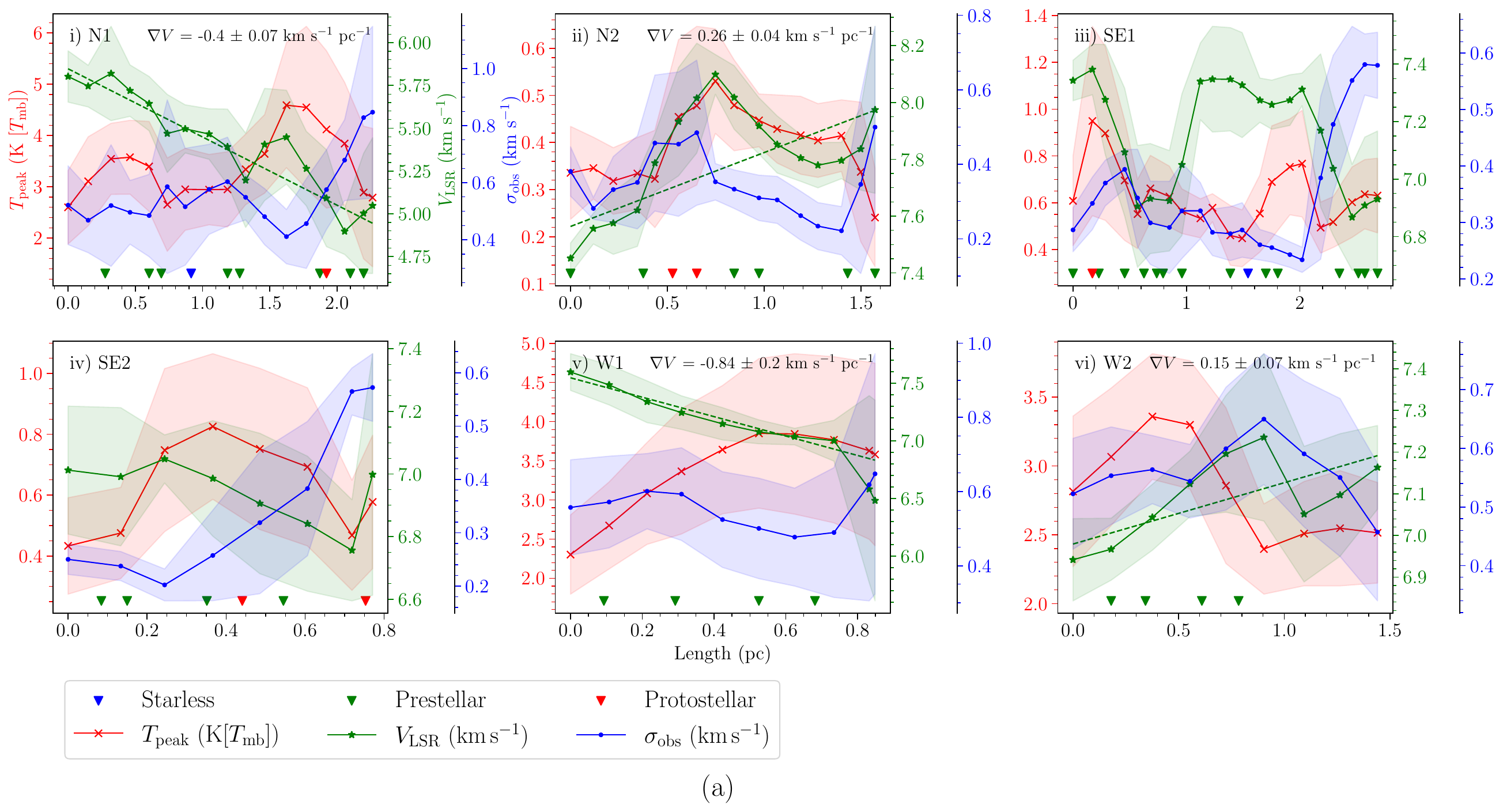}\\
    \includegraphics[width=\textwidth]{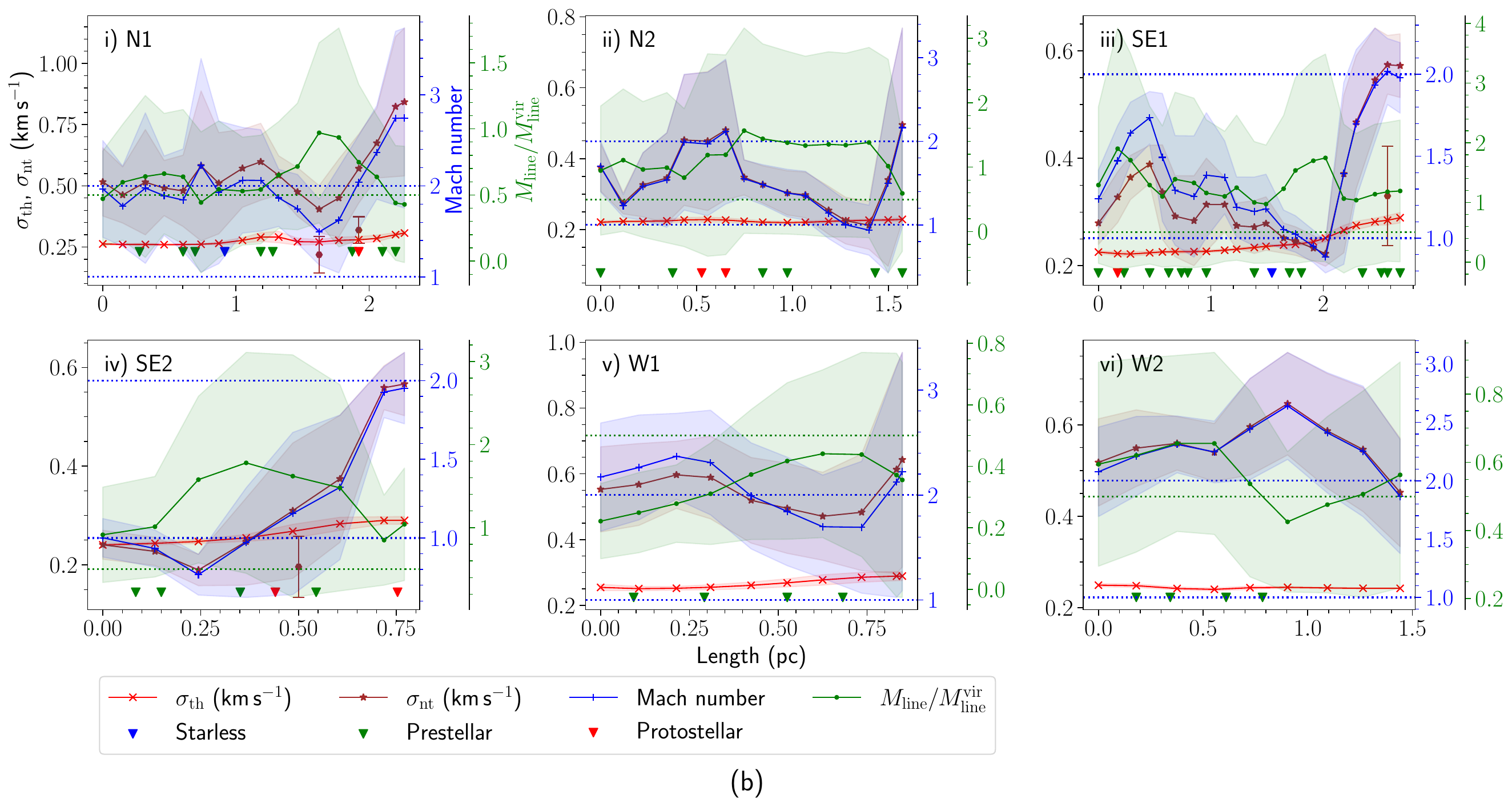}
    \caption{(a) Profiles of centroid velocity ($\vlsr$) (green $\star$), velocity dispersion ($\sigma_\text{obs}$) (blue $\cdot$), and peak temperature $T_\mathrm{peak}$ (red $\times$) for each identified filament, derived from single-component Gaussian fitting. Green dashed lines in sub-panel i), ii), v), and vi) indicate the best linear fit to the $\vlsr$ profile (velocity gradient, $\nabla V$), with its value and fitting error denoted on the upper right of the sub-panel. (b) Profiles of thermal (red $\times$), non-thermal (brown $\star$) velocity dispersions ($\sigma_\mathrm{th}$, $\sigma_\mathrm{nt}$), turbulent Mach number ($\sigma_\mathrm{nt}/\sigma_\mathrm{th}$) (blue $+$) and stability ratio ($M_\mathrm{line}/M_\mathrm{line}^\mathrm{vir}$) (green $\cdot$) of each identified filament. In all panels, the shaded region of respective colour represents the standard deviation of each data point, and inverted coloured triangles at the bottom of each panel indicate the positions of \emph{Herschel}-identified dense cores \citep{konyves2015}: blue for starless, green for prestellar, and red for protostellar. In panel group (b), blue dotted lines represent Mach numbers of 1 and 2, while green dotted lines represent $M_\mathrm{line}/M_\mathrm{line}^\mathrm{vir}=0.5$. Brown hexagon scatters in sub-panels i), iii), and iv) denote the $\sigma_\mathrm{nt}$ of N$_2$H$^+$ clumps (see Section~\ref{sec: clumps_analysis}). Axis labels shown in the top-left panels are common to all six sub-panels.}
    \label{fig: filament_T_vlsr_dv_profile}
\end{figure*}

Along each filament skeleton, we extracted profiles of peak temperature $T_\mathrm{peak}$, LSR velocity $\vlsr$, and observed velocity dispersion $\sigma_\mathrm{obs}$ by fitting single Gaussians to the spectra. These profiles are shown in Fig.~\ref{fig: filament_T_vlsr_dv_profile}a, with position ``0 pc" defined at the filament end farthest from central massive stars. Note that this only gives a rough estimate of the trends, since fitting a single Gaussian tends to overestimate the $\sigma_\mathrm{obs}$ of overlapping multi-Gaussian profiles and smooth out local variations in $\vlsr$. We derived the velocity gradient ($\nabla V$) of the filaments by fitting a straight line to the $\vlsr$ profile. The values with good fit are shown in the same figure.

We decomposed $\sigma_\mathrm{obs}$ into thermal ($\sigma_\mathrm{th}$) and non-thermal ($\sigma_\mathrm{nt}$) components, and derived the line mass ($M_\mathrm{line}$) and virial line mass ($M_\mathrm{line}^\mathrm{vir}$) profiles to evaluate stability against gravitational collapse ($M_\mathrm{line}^\mathrm{vir}/M_\mathrm{line}$). When deriving $\sigma_\mathrm{th}$, we assumed the kinetic temperature of the gas to be equal to the \emph{Herschel} dust temperature ($T_\mathrm{kin}=T_\mathrm{dust}$). The profiles are shown in Fig.~\ref{fig: filament_T_vlsr_dv_profile}b.

Because \emph{Herschel} dense cores are often found along the filaments \citep[e.g.,][]{andre_2010}, we marked their locations on the profile plots. A core is included if its perpendicular distance from the filament skeleton is less than one-quarter of the filament width.

The calculation procedures are detailed in Appendix~\ref{sec: filament_parameters_derived}. We report the mean and standard deviation of the parameters in Table~\ref{tab: filament_info}. Key results and interpretations based on these parameters are presented in the subsections below.

\subsubsection{Filament Width} \label{sec: filament_width} 
The filament widths\footnote{The reported width values are not deconvolved. Given the data resolutions, deconvolution would reduce widths by no more than 0.05\,pc.} derived in this study range from 0.21 to 0.36\,pc, which are resolved at the spatial resolutions of our data (0.05--0.1\,pc). These values are larger by a factor of 2--3 compared to those reported from dust continuum observations, which typically are found to be $\le$\,0.1\,pc \citep{menshchikov2010, arzoumanian2011, arzoumanian2019}. However, that earlier analysis assumed a distance of 260\,pc, roughly half of the 502\,pc adopted in our work. Scaling their reported widths to our assumed distance yields an expected width of $\sim$\,0.2\,pc. Nevertheless, our measured widths remain 1--2 times higher than this rescaled value. A similar discrepancy was reported by \citet{chung2021}, who found that filament widths derived from C$^{18}$O were systematically broader than those inferred from \emph{Herschel} dust continuum in the IC 5146 region.

Previous studies have reported different conclusions about filament widths. \citet{hacar2022} and \citet{panopoulou2022} found that measured widths often scale with angular resolution--up to ten times the beam size—suggesting a strong resolution dependence (see also Figure~1 in \citealt{panopoulou2022}). In contrast, \citet{andre2025a} showed that widths in NGC 6334 remain consistently $\sim$0.12 pc, even at JWST resolution.

Additionally, \citet{shimajiri2023} demonstrated in NGC 2024 that the derived width depends on the tracer used due to the differences in density regimes traced by various molecular lines.

We examined the effects of angular resolution and tracer choice on N1 and SE1 width measurements using our available data sets and found no significant discrepancies (see Appendix~\ref{sec: filament_length_and_width_derived} and Fig.~\ref{fig: filchap_width_profile}). However, the small sample size limits the statistical significance of this conclusion.

Finally, we find substantial variation in filament width along their lengths, consistent with previous reports in other star-forming regions \citep[e.g.,][]{suri2019, juvela2012}. This might be due to observational effects such as blending structures.

\subsubsection{Velocity dispersions} 
\label{subsubsec: filament_velocity_dispersions}
The thermal velocity dispersions ($\sigma_\mathrm{th}$), are similar across all filaments, ranging from 0.22 to 0.27\kms. $\sigma_\mathrm{th}$ is based on dust temperature ($T_\mathrm{dust}$), which ranges from 14 to 21\,K across the filaments, while its average value  is 18\,K. For comparison, the typical $T_\mathrm{dust}$ and $\sigma_\mathrm{th}$ of nearby clouds are 15\,$\pm$\,3\,K and is $\sim$\,0.22\kms\! \citep{arzoumanian2019}. Our slightly higher values likely reflect external heating from the \ion{H}{ii}\ region. 

The non-thermal velocity dispersions ($\sigma_\mathrm{nt}$) are $\sim$0.34\kms\! for H$^{13}$CO$^+$ detected filaments, and $\sim$0.55\kms\! for C$^{18}$O detected filaments. In all filaments, $\sigma_\mathrm{nt}>\sigma_\mathrm{th}$. 

We noticed an increase of $\sigma_\mathrm{obs}$, $\sigma_\mathrm{th}$ and $\sigma_\mathrm{nt}$ towards the right-hand side in Fig.~\ref{fig: filament_T_vlsr_dv_profile}a) in all filaments except W2. The variations in $\sigma_\mathrm{obs}$ and $\sigma_\mathrm{nt}$ also increase significantly. 

\subsubsection{$\vlsr$}
$\vlsr$ of all detected filaments are centred around 7\kms, except for N1 ($\sim$\,5.3\kms) and N2 ($\sim$\,7.8\kms). \citet{Kumar_2020} interpreted N2 as a filamentary structure likely swept up by the expanding \ion{H}{ii}\ bubble, potentially explaining its slightly shifted $\vlsr$. N1 exhibits the most distinct velocity, with a slight offset from the others. All filaments except SE1 and SE2 display velocity gradients, indicative of mass accretion within the HFS, with absolute values ranging from 0.15 to 0.84\kmspc. Among them, N1 and W1 exhibit particularly clear gradients. For comparison, reported values in other HFSs are generally lower, e.g., 0.09--0.31\kmspc\! in G321.93-0.01 \citep{maity2025}, 0.17--0.39\kmspc\! in W49A \citep{zhang2024}, 0.04--0.13\kmspc\! in G11P1-HFS \citep{bhadari2024} and 0.1--0.18\kmspc\! in G310 \citep{yang2023}. Therefore, our gradients appear somewhat higher, although we note that this comparison is not exhaustive, and other HFSs with larger values may exist.

The velocity components described in Section~\ref{subsec: molecular_cloud_distribution} and the \textsc{astrodendro} filaments represent different levels of the underlying gas structure. The 3, 5, 7, and 10\kms\! components correspond to large-scale bulk emission features, while \textsc{astrodendro} identifies smaller, velocity-coherent substructures within these components. Each filament falls entirely within a single velocity component (e.g., SE1, SE2, N2, W1, and W2 within the 7\kms\! component; N1 within the 5\kms\! component). No dendrogram structures are found in the 3 or 10\kms\! components. The two decompositions are therefore complementary, not contradictory: the velocity components describe the global kinematic environment, and the \textsc{astrodendro} structures reveal the coherent filaments embedded within those flows.

\begin{figure*}
    \centering
    \includegraphics[width=\textwidth]{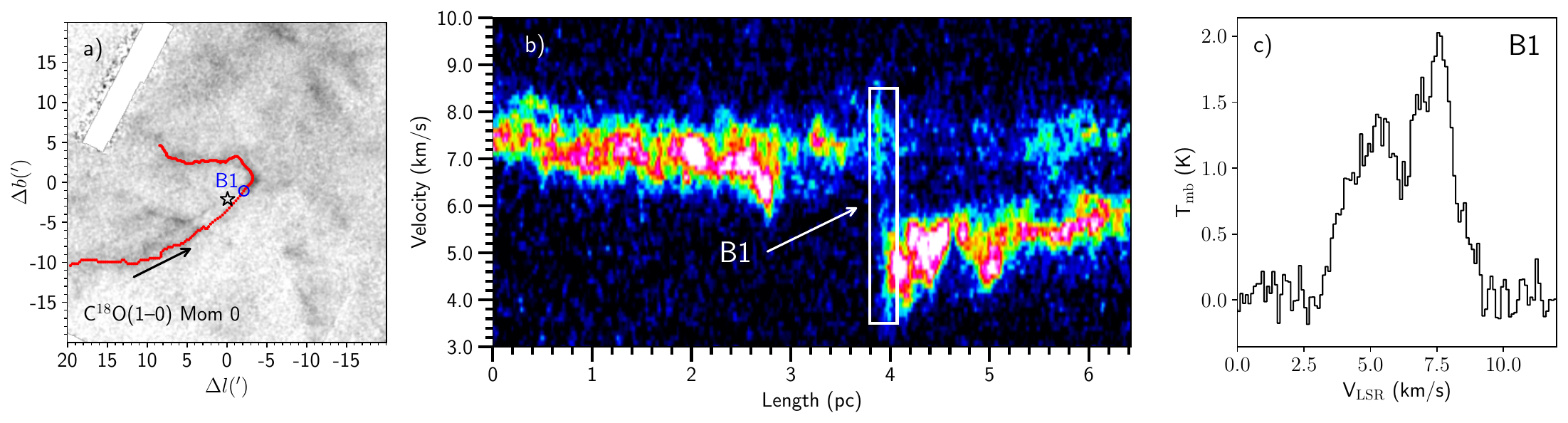}
    \caption{(a) C$^{18}$O moment-0 map, overlayed with the spatial path used to extract the position-velocity (PV) diagram of $\mathrm{C^{18}O}$ (red solid line), along with the position of B1 in position-position (PP) space. The black arrow indicates the direction of the PV cut from 0\,pc. 
    (b) The PV diagram. A bridge-like feature (B1) is observed at a path length of approximately 3.9\,pc. (c) Spectrum extracted from the circular region (in blue) marked in panel (a).}
    \label{fig: nro_c18o_pv_diagram}
\end{figure*}

\citet{Mallick_2013} proposed that SE1 and N1 formed independently and later merged, triggering the formation of dense cores--and eventually massive stars--at their junction. To explore this potential interaction, we constructed a position-velocity (PV) diagram connecting SE1 and N1 by hand, as shown in Fig.~\ref{fig: nro_c18o_pv_diagram}. The diagram reveals a bridge-like structure (B1) at a path length of approximately 3.9\,pc, suggesting possible physical interaction between the two filaments. To further investigate this feature, we also plotted the spectra at the bridge location (B1), shown in the same figure. At B1, the 5\kms\! and 7\kms\! components are both broad (FWHM=2.36 and 1.63\kms\! respectively) and partially overlapped. The implications of this kinematic bridge are discussed further in Section~\ref{subsec: hfs_origin}.

\subsubsection{$M_\mathrm{line}$ and $M\mathrm{_{line}^{virial}}$}
The mean line mass ($M_\mathrm{line}$) of all filaments lies in the range 51--96 M$_{\sun}$$\mathrm{\ pc}^{-1}$. The corresponding virial line mass ($M\mathrm{_{line}^{virial}}$) range is 172--187 M$_{\sun}$$\mathrm{\ pc}^{-1}$ with average value of 179 M$_{\sun}$$\mathrm{\ pc}^{-1}$ for the C$^{18}$O filaments; and 81--95 M$_{\sun}$$\mathrm{\ pc}^{-1}$ with average value of 88 M$_{\sun}$$\mathrm{\ pc}^{-1}$ for the H$^{13}$CO$^+$ filaments (see Appendix~\ref{sec: fil_dispersions_derivation} for calculation procedure). The ratios ($M_\mathrm{line}/M_\mathrm{line}^{\mathrm{virial}}$) therefore indicate that the H$^{13}$CO$^+$ filaments are transcritical ($\sim$1), i.e. close to gravitational equilibrium and prone to fragmentation, whereas the C$^{18}$O filaments are subcritical ($<1$) consistent with additional support (turbulent and/or magnetic) dominating over self-gravity \citep{ostriker1964, inutsuka1997}. The systematic difference between tracers is expected--H$^{13}$CO$^+$ preferentially traces denser filament skeletons, while C$^{18}$O traces more diffuse envelopes where velocity dispersion is higher, and density is lower. It is to be noted that all identified filaments in the sample are associated with \emph{Herschel}-identified dense cores, consistent with the picture of core formation proceeds within (near-)critical filaments \citep{andre_2010, arzoumanian2011, konyves2015}. Taken together, these results suggest that the denser H$^{13}$CO$^+$ filaments in the W40 cloud are near the  threshold for gravitational instability, while C$^{18}$O structures trace less bound, pressure-confined streams feeding the cores. 

\begin{table*}
\centering
\caption{Physical Properties of the filaments identified from molecular line observations. 
\label{tab: filament_info}}
\begin{tabular}{cccccccccc}
\hline
Fil. ID & 
Line &
$L$ &
$W$ &
$V_\mathrm{LSR}$ & 
$M_\mathrm{filament}$ &
$M_\mathrm{line}$ &
$M\mathrm{_{line}^{vir}}$ &
$\sigma_\mathrm{th}=c_s$ &
$\sigma_\mathrm{nt}$ \\
 &
 &
(pc) & 
(pc) &
(\kms) &  
($M_\odot$) &
($M_\odot \mathrm{\ pc}^{-1})$ &
($M_\odot \mathrm{\ pc}^{-1}$) &
(\kms) &
(\kms) \\
(1) &
(2) &
(3) &
(4) &
(5) &
(6) &
(7) &
(8) &
(9) &
(10) \\
\hline
N1  & C$^{18}$O(1--0)      & 2.25\,($\pm$\,0.08) & 0.29\,($\pm$\,0.13) & 5.41\,($\pm$\,0.28) & 237\,($\pm$\,60)           & 80\,($\pm$\,32) & 187\,($\pm$\,93)           & 0.27 & 0.55\,($\pm$\,0.10) \\
N2  & H$^{13}$CO$^+$(1--0) & 1.57\,($\pm$\,0.14) & 0.21\,($\pm$\,0.19) & 7.81\,($\pm$\,0.18) & 143\,($\pm$\,39)           & 77\,($\pm$\,82) & \phantom{1}81\,($\pm$\,41) & 0.22 & 0.34\,($\pm$\,0.11) \\
SE1 & H$^{13}$CO$^+$(1--0) & 2.68\,($\pm$\,0.14) & 0.27\,($\pm$\,0.23) & 7.16\,($\pm$\,0.18) & 277\,($\pm$\,67)           & 94\,($\pm$\,93) & \phantom{1}87\,($\pm$\,21) & 0.24 & 0.34\,($\pm$\,0.06) \\
SE2 & H$^{13}$CO$^+$(1--0) & 0.77\,($\pm$\,0.14) & 0.24\,($\pm$\,0.06) & 6.94\,($\pm$\,0.09) & \phantom{1}74\,($\pm$\,30) & 96\,($\pm$\,63) & \phantom{1}95\,($\pm$\,23) & 0.26 & 0.34\,($\pm$\,0.07) \\
W1  & C$^{18}$O(1--0)      & 0.85\,($\pm$\,0.08) & 0.21\,($\pm$\,0.04) & 7.10\,($\pm$\,0.34) & \phantom{1}47\,($\pm$\,17) & 51\,($\pm$\,25) & 177\,($\pm$\,79)           & 0.27 & 0.55\,($\pm$\,0.17) \\
W2  & C$^{18}$O(1--0)      & 1.44\,($\pm$\,0.08) & 0.36\,($\pm$\,0.10) & 7.09\,($\pm$\,0.09) & 142\,($\pm$\,50)           & 80\,($\pm$\,32) & 172\,($\pm$\,52)           & 0.24 & 0.55\,($\pm$\,0.10) \\
\hline
\end{tabular}
\medskip
\begin{minipage}{0.97\textwidth}
\textit{Column descriptions}: 
(1) Filament designation; 
(2) Molecular line used to identify the filament; 
(3) Filament length; 
(4) Filament width; 
(5) Mean centroid LSR velocity; 
(6) Total filament mass; (7) Mean line mass of the filament; 
(8) Mean virial line mass, representing gravitational equilibrium; (9) Mean thermal velocity dispersion  (isothermal sound speed); 
(10) Mean non-thermal velocity dispersion. 
Uncertainties are estimated as follows: 
$W$ from the standard deviation of the \texttt{FilChap} width profile (Figure~\ref{fig: filchap_width_profile}); 
$L$ from distance and beam-size uncertainties; 
$M_\mathrm{filament}$ as detailed in Section~\ref{sec: filament_parameters_derived}; 
and all other parameters from the standard deviation of their profiles (Figure~\ref{fig: filament_T_vlsr_dv_profile}). 
Uncertainties of $\sigma_\mathrm{th}$ is negligible (not shown).
\end{minipage}
\end{table*}

\subsection{Clump Identification and Analysis} \label{sec: clumps_analysis}
\begin{figure}
    \centering
    \includegraphics[width=\columnwidth]{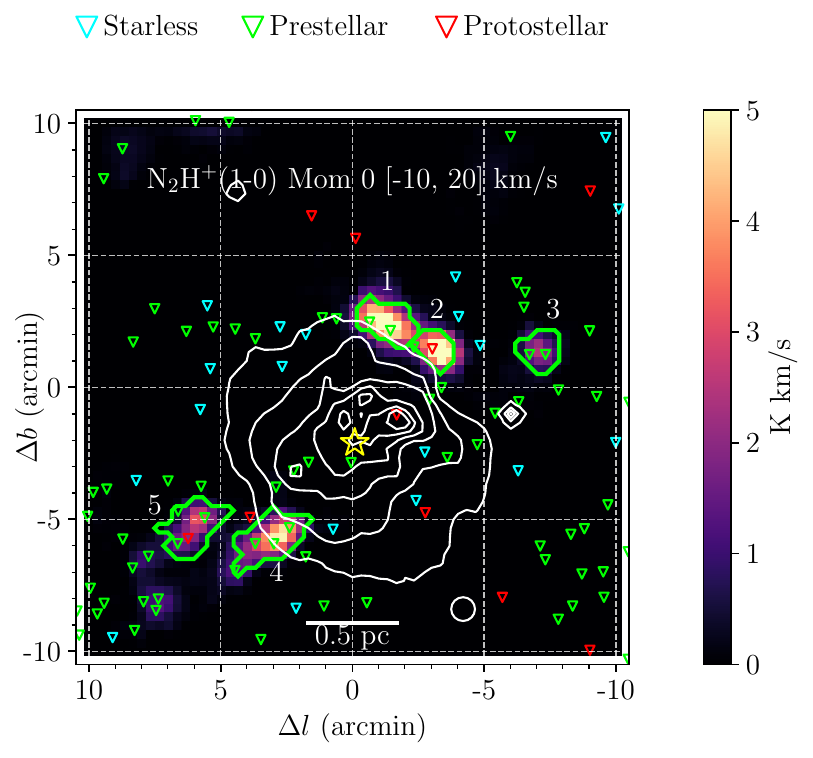}
    \caption{Identified clumps overlaid on the N$_2$H$^+$ moment-0 map integrated over the velocity range of$-$10 to 20\kms. The spatial distribution of \emph{Herschel} dense cores \citep{konyves2015} is also overlaid. The white contours are drawn similarly to Fig.~\ref{fig: w40_complex}a). The spatial distribution of Clump 1, 2, and 4 suggests that these clumps may result primarily from local instabilities triggered by stellar feedback, rather than from large-scale gravitational infall alone.}
    \label{fig: herschel_dense_cores_on_trao_n2h+_mom0}
\end{figure}

N$_2$H$^+$ traces denser gas compared to C$^{18}$O, and therefore reveals compact, locally denser regions along the filaments (see Figure\,\ref{fig: w40_moment0_maps_rgb}i), which coincide with \emph{Herschel} identified prestellar and protostellar cores (see Fig.~\ref{fig: w40_complex}d). We identified these clumps as velocity-coherent structures using \textsc{astrodendro} package on the TRAO N$_2$H$^+$ data cube and derived their physical parameters. A detailed description of the identification method and derivation of physical parameters is provided in Appendix~\ref{sec: clump_identification}. We identified a total of 5 clumps. The identified clumps are shown in Fig.~\ref{fig: herschel_dense_cores_on_trao_n2h+_mom0}.  The derived clump properties are summarized in Table~\ref{tab: clumps_info}. 

Clump 3 is unique among the identified clumps, being isolated and not clearly associated with any filament
. It also exhibits a nearly constant C$^{18}$O integrated intensity ($I$\,$\sim$\,4\Kkms), suggesting a distinct environment compared to other clumps. Its lower dust temperature ($T_\mathrm{dust}$$\sim\,$16\,K) is consistent with CO depletion due to freeze-out onto dust grains \citep[e.g.,][]{caselli1999,bacmann2002}. 

Initially, assuming the gas kinetic temperature ($T_\mathrm{k}$) equals the excitation temperature, $T_\mathrm{ex}$ ($T_\mathrm{k}=T_\mathrm{ex}$), all clumps appear supersonic ($\sigma_\mathrm{nt}/c_s>2$). However, N$_2$H$^+$ typically shows sub-thermal excitation, with $T_\mathrm{ex}$ values 4--5 times lower than $T_\mathrm{dust}$, a behaviour also observed in Class 0 YSOs \citep{redaelli2020}. Thus, the assumption  $T_\mathrm{k}=T_\mathrm{ex}$ underestimates thermal velocity dispersion ($\sigma_\mathrm{th}$) by a factor of $\sim$2. Under the more realistic assumption of $T_\mathrm{k}=T_\mathrm{dust}$, the clumps exhibit subsonic to transonic turbulence. 

Clumps 1 and 5 have a virial parameter $\alpha_\mathrm{vir}<1$, indicating they are gravitationally bound and likely undergoing collapse. Clump 4, with $1<\alpha_\mathrm{vir}<2$, is near virial equilibrium. Clumps 2 and 3, with $\alpha_\mathrm{vir}>2$, appear to be gravitationally unbound and dominated by kinetic energy. For Clump 3, turbulence may originate from the ambient medium, whereas turbulence in Clumps 2 and 4 may be influenced by feedback from massive stars due to their proximity to the \ion{H}{ii}\ region. Interestingly, all identified clumps spatially coincide with at least one \emph{Herschel} prestellar core. We note that all the clumps have inner sources (e.g., dense cores), which could also influence their kinematics and stability.

To compare clump-scale non-thermal dispersions ($\sigma_\mathrm{nt}$) with their parent filament-scale values (Section~\ref{sec: filament_analysis}), we also plot the $\sigma_\mathrm{nt}$ of Clump 1, 2, 4 and 5 in Fig.~\ref{fig: filament_T_vlsr_dv_profile}. We find in all cases a lower $\sigma_\mathrm{nt}$ values of the clumps than those of the parental filaments. This trend suggests turbulence dissipation during the transition from filament 
to clump scales \citep{offner2022}.

{\setlength{\tabcolsep}{4pt}
\begin{table*}
\centering
\caption{Physical Properties of the dense clumps identified from N$_2$H$^+$(1--0) observations.}
\label{tab: clumps_info}
\begin{tabular}{cccccccccccc}
\hline
ID & 
$a$ &
$b$ &
$T_\mathrm{ex}$ &
$T_\mathrm{dust}$ & 
$N\mathrm{(N_2H^+)}$ &
$V_\mathrm{LSR}$ &
$M_\mathrm{clump}$ &
$c_s=\sigma\mathrm{_t}$ &
$\sigma\mathrm{_{nt}}$ &    
$\sigma\mathrm{_{nt}/\sigma\mathrm{_t}}$ &
$\alpha_\mathrm{vir}$ \\
 &
(pc) &
(pc) &
(K) &
(K) & 
($\times10^{12}$\,$\mathrm{cm}^{-2}$) &
(\kms) &
($M_{\sun}$) &
(\kms) &
(\kms) &
 &
  \\
(1) &
(2) &
(3) &
(4) &
(5) &
(6) &
(7) &
(8) &
(9) &
(10) &
(11) &
(12) \\
\hline
1 & 0.38 & 0.22 & 4.3\,($\pm$\,0.3) & 21.5\,($\pm$\,1.7) &            17.7\,($\pm$\,10.1) & 5.30\,($\pm$\,0.12) & 36\,($\pm$\,21) & 0.27\,($\pm$\,0.02) & 0.22\,($\pm$\,0.08) & 0.79\,($\pm$\,0.40) & 0.58\,($\pm$0.37)  \\
2 & 0.26 & 0.19 & 5.3\,($\pm$\,1.0) & 20.6\,($\pm$\,1.3) &  \phantom{1}8.8\,($\pm$\,6.6) & 4.70\,($\pm$\,0.20) &  11\,($\pm$\,8) & 0.27\,($\pm$\,0.01) & 0.34\,($\pm$\,0.07) & 1.26\,($\pm$\,0.39) & 2.26\,($\pm$1.80)  \\
3 & 0.25 & 0.24 & 3.5\,($\pm$\,0.5) & 16.2\,($\pm$\,0.9) &  \phantom{1}7.6\,($\pm$\,5.6) & 6.22\,($\pm$\,0.18) &  11\,($\pm$\,8) & 0.24\,($\pm$\,0.01) & 0.35\,($\pm$\,0.05) & 1.45\,($\pm$\,0.32) & 2.28\,($\pm$1.72)  \\
4 & 0.48 & 0.25 & 3.7\,($\pm$\,0.4) & 22.5\,($\pm$\,1.4) &  \phantom{1}9.3\,($\pm$\,7.3) & 7.04\,($\pm$\,0.13) & 26\,($\pm$\,21) & 0.28\,($\pm$\,0.01) & 0.36\,($\pm$\,0.10) & 1.28\,($\pm$\,0.54) & 1.61\,($\pm$1.38)  \\
5 & 0.41 & 0.27 & 3.9\,($\pm$\,0.6) & 19.9\,($\pm$\,2.4) &  \phantom{1}9.1\,($\pm$\,5.6) & 6.91\,($\pm$\,0.17) & 23\,($\pm$\,15) & 0.27\,($\pm$\,0.02) & 0.22\,($\pm$\,0.07) & 0.81\,($\pm$\,0.40) & 0.97\,($\pm$0.66)  \\
\hline
\end{tabular}
\medskip
\begin{minipage}{0.97\textwidth}
\textit{Column descriptions}: 
(1) Clump ID;
(2) Major axis;
(3) Minor axis;
(4) LTE-derived N$_2$H$^+$ excitation temperature; 
(5) Dust temperature obtained from {\it Herschel} data \citep{konyves2015};
(6) Average N$_2$H$^+$ column density of the clumps; 
(7) Average LSR velocity of the clumps; 
(8) Clump mass;
(9) Thermal velocity dispersion (i.e., isothermal sound speed);
(10) Non-thermal velocity dispersion;
(11) Turbulent Mach number;
(12) Virial parameter of the clump.
The derivations of the parameters and their uncertainties are detailed in Section~\ref{sec: clump_parameters_derived}.
\end{minipage}
\end{table*}
}

\section{Discussion}
\label{sec:discussion}

\subsection{Massive Star Feedback and Triggered Star Formation} 
\label{sec: triggered_star_formation}
The increasing trends of $\sigma_\mathrm{th}$ and $\sigma_\mathrm{nt}$ along the filaments as they converge towards the \ion{H}{ii}\ region (see Section~\ref{subsubsec: filament_velocity_dispersions}) can be interpreted as the effect from massive star feedback--in the form of increasing temperature and turbulence. Similar findings have been reported in numerical simulations \citep[e.g.,][]{suin2025} and observations \citep[e.g.,][]{yang2023}.

The massive star feedback appears to be positive in W40, with dense N$_2$H$^+$ clumps forming at the edge of the \ion{H}{ii}\ region: Clump 1, 2 are located on the edge of the 887.5 MHz contour, while Clump 4 coincides with it. The density might be enhanced by the collection of the material in the ionization-front/shock-front interface due to the expansion of the \ion{H}{ii}\ region. At the same time, shielding from the PDR prevents N$_2$H$^+$ from being destroyed by free electrons, and the elevated $T_\mathrm{dust}$ might further boost its abundance \citep{yu2018}.

Previous studies have proposed that W40 is undergoing multiple stages of star formation \citep[e.g.,][]{maury2011, pirogov2015, shimoikura2015}, with particular attention to the western ring-like dust structure near IRS 5 (see Fig.~\ref{fig: w40_complex}d for IRS 5 position). \citet{maury2011} noted that this ring is predominantly populated by Class 0 protostars, in contrast to the central cluster, which is rich in more evolved Class II and III objects. This spatial segregation suggests that the ring is younger compared to regions hosting Class II and III objects, and is potentially created by the expansion of the central \ion{H}{ii}\ region. Supporting this view, \citet{shimoikura2015} detected multiple $^{12}$CO(3--2) outflow lobes within the ring, many of which are associated with HCO$^+$(4--3) clumps, providing further evidence of ongoing, active star formation. \citet{pirogov2015} proposed a collect-and-collapse scenario to explain the formation of low-mass clumps along the ring, showing that the observed structures could be reproduced under conditions of a high initial cloud density ($\ge10^5\,$cm$^{-3}$) and a strong ionizing flux ($3\times10^{46}$\,photons\,s$^{-1}$). Consistent with this, we detect H$^{13}$CN emission—indicative of dense gas ($10^5$\,cm$^{-3}$)--coinciding with the ring structure (see Fig.~\ref{fig: w40_moment0_maps_rgb}f),  further supporting the idea of triggered star formation scenario in the ring.

In this work, we propose a second site of potentially triggered star formation: the waist of the bipolar bubble. This interpretation is based on the detection of clumpy C$_2$H emission and corresponding infrared features.

We detect localized, clumpy C$_2$H emission at $\sim$\,10\kms\! along the waist of the bipolar structure (see Fig.~\ref{fig: w40_moment0_maps_rgb}k), suggesting that this region is actively influenced by feedback processes. C$_2$H is a dense gas tracer commonly observed both in the early stages of high-mass star formation \citep{beuther2008,martinez2024} and in PDRs, where the gas is shaped by radiation and stellar winds. Its presence here may therefore reflect either the onset of gravitational collapse or compression induced by radiative feedback, or the combination of both. The emission coincide with two previously reported outflow candidates (R3 and R4; \citealt{shimoikura2015}) based on the excess emission in $^{12}$CO(3--2) red wing ($11.7\kms\!\le\vlsr\le14.5\kms$). 

Moreover, the regions with C$_2$H emission coincide with ``elephant trunk" or bright-rimmed cloud (BRC) features in infrared images. Such morphological structures are often associated with scenarios where ionizing radiation interacts with pre-existing dense clumps--in particular via radiation-driven implosion (RDI)--which can lead to compression and possibly trigger star formation \citep[e.g.,][]{sugitani1989, sugitani1991, sugitani1994, bisbas2011}. However, similar pillar-like features can also arise from alternative mechanisms, such as dynamical instabilities \citep{tremblin2013}.

Together, these results support the idea that feedback from massive stars in W40 is not only responsible for the well-known ring structure, but may also be inducing a newer episode of star formation along the waist of the expanding \ion{H}{ii}\ region. Dense clumps surrounding the waist of other bipolar \ion{H}{ii}\ regions are also observed, and interpreted as star formation triggered by the expansion of central \ion{H}{ii}\ region (e.g., G338.93-00.06, G316.804-400.05 \cite[][]{samal2018}; and  G010.32-00.15 \cite[][]{deharveng2015}). 

\subsection{The Origin of HFS} \label{subsec: hfs_origin}
\citet{Kumar_2020} identified the W40 complex as an evolved stage of HFS, where the current location of the cluster of massive stars represents the remnant of such a hub where mass had accumulated in the past. This is contrast to an early stage of HFS, where mass accretion is observed along the filaments seen by velocity gradient of dense gas tracer, and is enhanced towards the hub, such as in the case of MonR2 \citep{trevino-morales2019}. To investigate the formation of the HFS in the W40 complex, we consider the mechanisms proposed for flow-driven filament formation, as discussed by \citet{Kumar_2020}. Such filaments can arise from intra-molecular cloud velocity dispersion or may be externally compressed by feedback processes such as expanding shells, stellar winds, \ion{H}{ii}\ region expansions, or supernova shocks. 

Numerical simulations also demonstrate that HFSs can form through cloud-cloud collisions (CCC) \citep{inoue2013,fukui2018,maity2024,inoue2018}. Observational signatures of CCC include: (1) bridge-like features in position-velocity (PV) diagrams connecting two velocity components, and (2) complementary spatial structures in position-position (PP) space (see review by \citealt{fukui2021}). In our analysis, we detected a bridge-like feature in the C$^{18}$O PV diagram connecting SE1 and N1 (Fig.~\ref{fig: nro_c18o_pv_diagram}). Additionally, two velocity features ($\vlsr\!\sim$\,5.5 and $\sim$\,7.5\kms) are seen overlapping in the PP space, north-eastern direction of the massive stars (see Fig.~\ref{fig: w40_moment0_maps_rgb}d). This overlapping feature is also seen in the PV diagram (Fig.~\ref{fig: nro_c18o_pv_diagram}b), between 5.4 to 6.4\,pc.

This location also coincides with compact high-density gas traced by CS(5--4) and C$^{34}$S(2--1) observed by \cite{Pirogov_2013}, as well as with peaks in 1280 and 610\,MHz continuum emission (see their Figures 2 and 6). These features are situated near the ionizing massive stars. Additionally, \cite{rumble2016} reported elevated dust temperatures at the interface based on JCMT observations, which further supports the presence of compression or heating.

According to \cite{dewangan2022}, clusters of YSOs and massive stars are commonly found at the intersection zones of colliding clouds or within the shock-compressed layers--consistent with what we observe in W40. Furthermore, simulations by \citet{gomez2014}, modelling the collision of two oppositely directed warm gas streams at a relative velocity of 18.4\kms, produce filamentary structures and PV diagrams resembling our observations after 26.56\,Myr into the simulation (see their Figure~4). Thus, filaments SE1 and N1 are probably the parental structure from which the central massive stars formed, and later disrupted by the expanding \ion{H}{ii}\ region. We note that N1, due to its high curvature and its vicinity to the \ion{H}{ii}\ region, could alternatively be dense material compressed by expansion of \ion{H}{ii}\ region. However, the fact that velocity gradient is observed there favours the former interpretation. 

Taken together, these observational and theoretical clues support the scenario where the dense filaments (N1 and SE1) in the W40 cloud are the result of shock-compressed layers produced by CCC. The resulting increase in local density likely triggered the formation of the massive stars. A similar formation mechanism has been proposed for other regions (e.g., W33 \citep{kohno2018,dewangan2020,zhou2023}; G31.41+0.31 \citep{beltran2022}; G013.313+0.193 \citep{berdikhan2025}). A search of complementary structures in PP space using wide field CO survey (e.g., The Milky Way Imaging Scroll Painting (MWISP; \citealt{su2019}) will be essential to further support this hypothesis. 

\section{Summary and Conclusions}
\label{sec:conclusion}
We have carried out a comprehensive multi-wavelength investigation of the HFS in the massive SF region W40, combining new TRAO molecular line observations with ancillary data sets. Our main findings are as follows:
\begin{enumerate}
    \item A pronounced deficit of emission around $\sim$\,7\kms\! is seen in most molecular line observations, accompanied by a blueshifted dip in the HCN profile of the 7\kms\! component. This points to the presence of a cold, foreground 7\kms\! gas component 
    that absorbs emission from a warmer background cloud, likely radiatively heated by the OB association, thereby producing the observed self-absorption features.
    \item Six velocity-coherent filaments are identified in C$^{18}$O and H$^{13}$CO$^+$ data, four of which converge towards the O9.5V star IRS 1A South, forming the hub. 
    \item Two filaments exhibit negative velocity gradients ($|\nabla V|\!=\!0.4\!-\!0.84\kmspc$) towards the hub, indicative of residual mass inflow. 
    \item Non-thermal velocity dispersions increase in filaments closer to the \ion{H}{ii} region, suggesting feedback-induced turbulence.  
    \item Five dense clumps are detected in N$_2$H$^+$ emission, showing subsonic to transonic turbulence, in contrast to the supersonic motions in their parental filaments--indicating turbulence dissipation at smaller scales. One clump, unassociated with any filament, exhibits the lowest mass and dust temperature. 
    \item A bridge-like feature in position-velocity space, coinciding with radio continuum peaks and dense gas condensations at the interface between two filaments (N1 and SE1)
    , suggests a past collision between two clouds. Such an event would have created a turbulent, shocked environment suitable for the formation of the HFS and massive stars.
    \item The overall morphology and kinematics indicate that the W40 complex represents a late-stage (Stage IV) HFS, where stellar feedback dominates. The observed structures align with scenarios involving 
    CCC followed by feedback-driven dispersal of the parental gas.  
\end{enumerate}
The W40 complex thus offers a valuable case study for understanding both the early assembly of massive SF hubs and their subsequent evolution under the influence of stellar feedback. High-resolution, wide-field observations in multiple rotational transitions of different density and temperature tracers combining with numerical simulations and chemical modellings will be critical for further constraining the cloud's kinematics, internal dynamics, and chemical evolution.

\section*{Acknowledgements}
We sincerely thank the referee for the constructive comments and suggestions, which have significantly improved the quality of the paper. This study relies on observations conducted using the TRAO 14-m radio telescope, which is managed by the Korea Astronomy and Space Science Institute. We extend our thanks to the staff and personnel of the TRAO 14-m radio telescope for their support throughout the observation period, as well as for their contribution to the data acquisition. This work is supported by the Fundamental Fund of Thailand Science Research and Innovation (TSRI) through the National Astronomical Research Institute of Thailand (Public Organization) (FFB680072/0269). N.S. and A.M.J acknowledges support by the CRC1601 (SFB 1601 sub-project B2) funded by the DFG (German Research Foundation) – 500700252. LEP is supported by the Russian Science Foundation grant 24-12-00153.

\subsection*{Data availability}
The TRAO 14-m telescope data underlying this article will be shared on reasonable request to the corresponding author. The Nobeyama 45-m telescope data sets were retrieved from the JVO portal operated by ADC/NAOJ (https://jvo.nao.ac.jp/portal/nobeyama/sfp.do). The JCMT HARP $^{12}$CO(3--2) data were obtained from the https://doi.org/10.11570/16.0006 \citep{rumble2016}. The IRAM 30-m telescope data were provided through private communication with Y. Shimajiri. The \emph{Spitzer} 8\micron\ SEIP image was obtained from the NASA/IPAC Infrared Science Archive (IRSA) (https://irsa.ipac.caltech.edu/data/SPITZER/Enhanced/SEIP/), and the H$\alpha$ image from the UKST SuperCOSMOS H$\alpha$ Survey (http://www-wfau.roe.ac.uk/sss/halpha/hapixel.html). The HGBS products, including column density and dust temperature maps, were retrieved from the HGBS data portal (http://www.herschel.fr/cea/gouldbelt/en/). The 887.5\,MHz radio continuum image is from the RACS Data Release 1 (https://research.csiro.au/racs/), and the 850\,µm image was obtained from the JCMT Gould Belt Survey DR3 archive (https://doi.org/10.11570/18.0005;\citealt{kirk2018}).

%%%%%%%%%%%%%%%%%%%% REFERENCES %%%%%%%%%%%%%%%%%%

\bibliographystyle{mnras}
\bibliography{main} % if your bibtex file is called example.bib
%%%%%%%%%%%%%%%%%%%%%%%%%%%%%%%%%%%%%%%%%%%%%%%%%%

%%%%%%%%%%%%%%%%% APPENDICES %%%%%%%%%%%%%%%%%%%%%
\appendix
\FloatBarrier
\section{C$^{18}$O spectra inconsistency}
\begin{figure}
    \centering
    \includegraphics[width=\columnwidth]{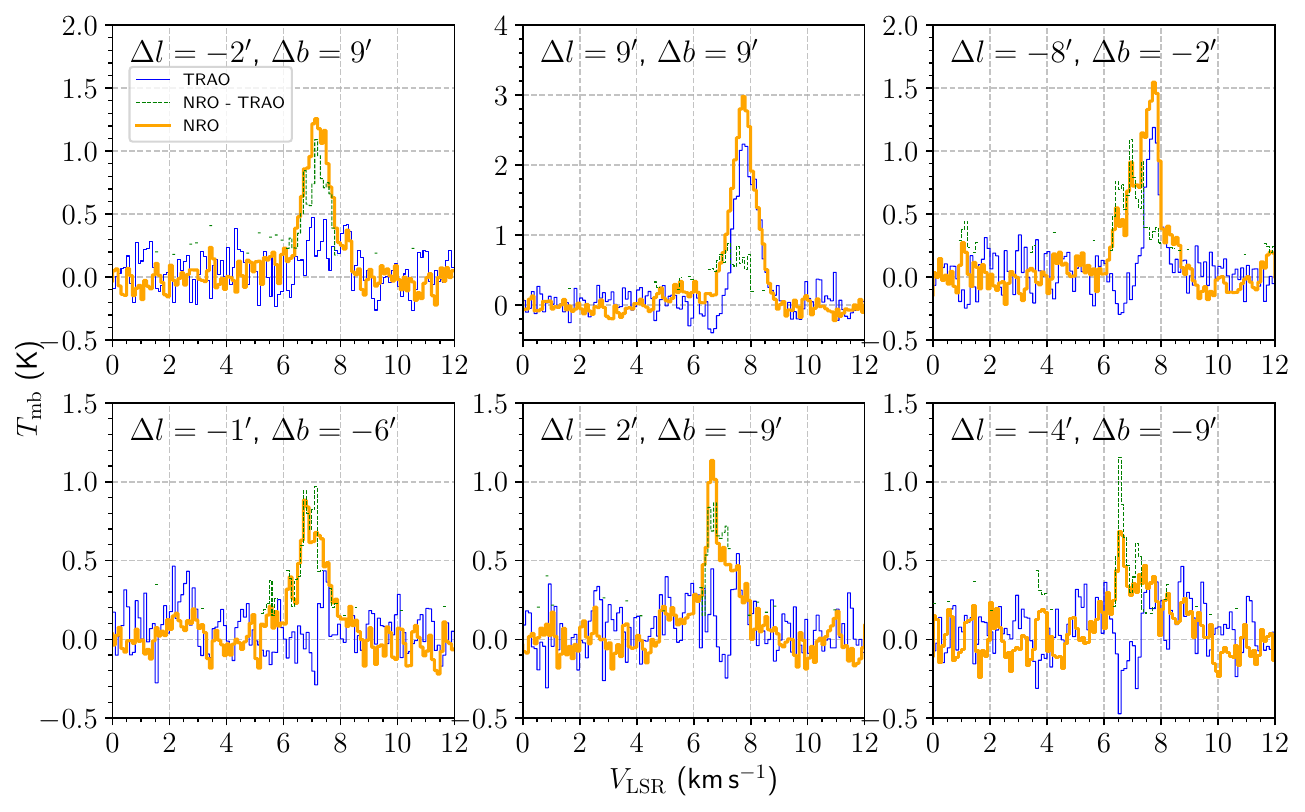}
    \caption{Comparison of C$^{18}$O spectra at selected positions between NRO (orange) and TRAO (blue) observations. Prior to comparison, both data sets were convolved to a common spatial and spectral resolutions. Each spectrum represents the average within a circular region of 1\arcmin diameter centred at the indicated offset coordinates. The dashed green line shows the difference between the two spectra.}
    \label{fig: c18o_nro_trao}
\end{figure}
\FloatBarrier   
\section{Molecular Average Spectra} 
\label{sec: avg_spectra}
Directly averaging a data cube that contains multiple velocity components can suppress weak, spatially compact features in the resulting spectra. To mitigate this, we constructed channel maps with a width of 2\kms\! covering the velocity range from $-$3 to 15\kms. For each channel map, we applied a threshold mask of 1\,K\kms\! (note: for C$_2$H, we adopted a lower threshold of 0.3\,K\kms\! due to its weaker emission) and then computed the mean spectrum within the masked area. The final averaged spectrum was then obtained by averaging over all these channel-based spectra.  

\begin{figure}
    \centering
    \includegraphics[width=\columnwidth]{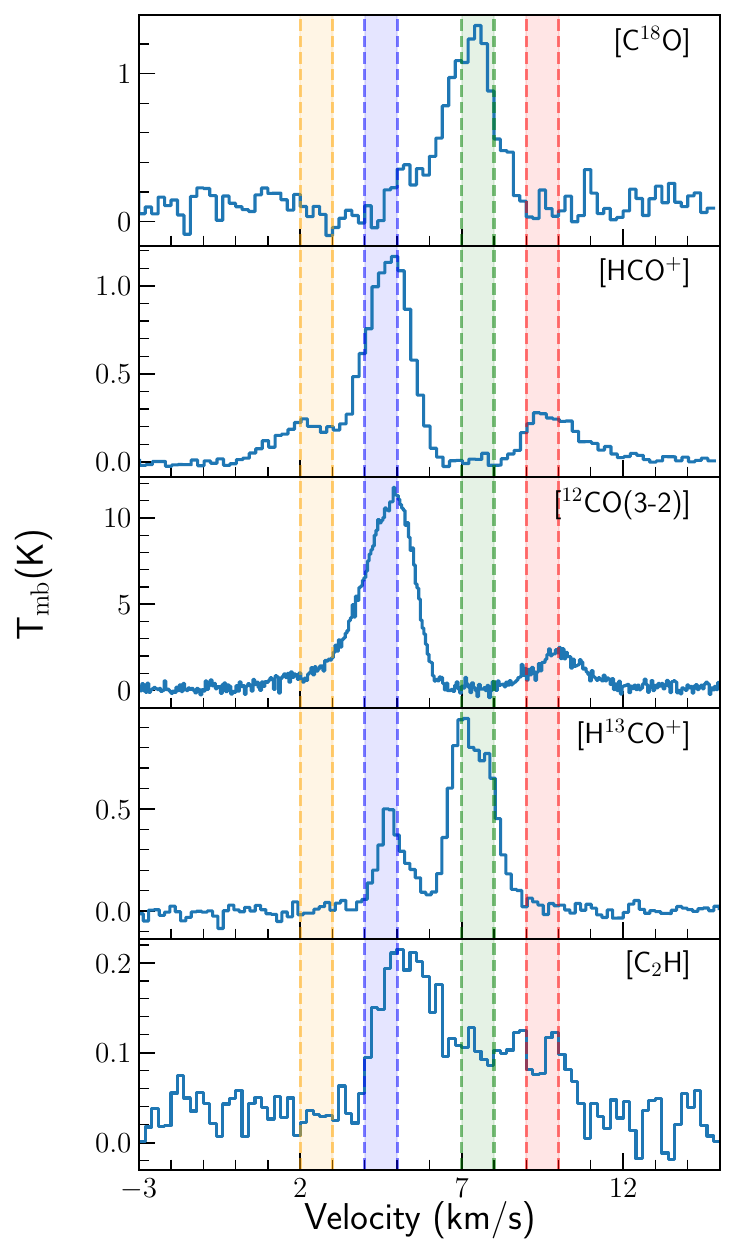}
    \caption{Averaged spectra of five selected molecular lines used to identify distinct velocity components in the W40 cloud 
    . The shaded regions indicate the four velocity components identified by visual inspection. Lines without transition labels correspond to the J=1--0 transition. Spectra of C$^{18}$O, HCO$^+$, and C$_2$H are smoothed to 0.2\kms\! before averaging.}
    \label{fig: avg_spectra}
\end{figure}

Fig.~\ref{fig: avg_spectra} displays the averaged spectra for a selection of molecular lines. We excluded transitions with closely spaced hyperfine lines (e.g., HCN) and optically thick lines with high abundance (e.g., $^{12}$CO) because they complicate the identification of distinct velocity components. 

\section{Filament and Clump Identification Using \textsc{astrodendro}} 
\label{sec: astrodendro_explained}
The \textsc{astrodendro} package decomposes a data cube into a hierarchical structure consisting of three components: leaves, which are the smallest structures that cannot be further subdivided; branches, which group together multiple leaves or smaller branches; and the trunk, which represents the largest connected structure. We use this framework to identify both filamentary and clump-like structures, applying different criteria depending on the morphology and kinematics of interest. We refer the readers to \citet{rosolowsky2008} for a detailed description of the algorithm, and \href{https://dendrograms.readthedocs.io/en/stable/algorithm.html}{\textsc{astrodendro} documentation} for the definitions of the parameters.

\subsection{Filaments} \label{sec: filament_identification}
Filaments were extracted by applying \textsc{astrodendro} to the NRO C$^{18}$O and IRAM H$^{13}$CO$^+$ data cubes. Candidate filaments were required to have an aspect ratio of the major to minor axis of a detected structure greater than 3, following \citet{arzoumanian2019}. When multiple leaves were identified within a single filamentary branch, only the longest leaf was retained to avoid counting smaller embedded substructures as independent filaments. Filaments associated with Serpens South were excluded from this analysis.

For the dendrogram construction, different parameter sets were adopted for the two tracers:

C$^{18}$O: \texttt{min\_value} = 3$\sigma$, \texttt{min\_delta} = 1$\sigma$, \texttt{min\_npix} = 400.

H$^{13}$CO$^+$: \texttt{min\_value} = 5$\sigma$, \texttt{min\_delta} = 1$\sigma$, \texttt{min\_npix} = 400.

Here, $\sigma$ denotes the rms noise of emission-free channels of each cube. 

\subsection{Clumps} 
\label{sec: clump_identification}
Clumps were identified as velocity-coherent structures in the TRAO N$_2$H$^+$ cube using \textsc{astrodendro}. The analysis was restricted to the velocity range [9, 15]~\kms, isolating the ($F_1\!=\!1\!-\!1, F\!=\!1\!-\!0, 2\!-\!2, 0\!-\!1$) hyperfine lines for cleaner detection.

Three key dendrogram parameters were adopted: \texttt{min\_value}, \texttt{min\_delta}, and \texttt{min\_npix}, set to 3$\sigma$, 3$\sigma$, and 27 pixels, respectively. Similarly, $\sigma$ is the rms noise of emission-free channels of the cube. The \texttt{min\_npix} value corresponds to roughly three times the beam area in pixels.

From this procedure, \textsc{astrodendro} identified 6 structures, of which 5 spatially isolated clumps were retained for further analysis.

\section{Physical Parameters of the Filaments} 
\label{sec: filament_parameters_derived}
\subsection{Length and Width} 
\label{sec: filament_length_and_width_derived}
\texttt{FilChaP} code computes radial intensity profiles perpendicular to the filament skeleton. Instead of averaging over the entire filament length, the code calculates these profiles in segments. This localized approach allows the fitting range for each profile to be set dynamically based on the local intensity minima surrounding the filament peak. We refer readers to \cite{suri2019} for an extensive description of \texttt{FilChaP}. \texttt{FilChaP} provides several methods for deriving filament widths, including Gaussian fits, two Plummer-like fits ($p$=2 and $p$=4), and the FWHM computed from the second moment of the intensity distribution. In the present analysis, we adopted the Gaussian fit-derived widths as our primary measurements.

The input file is a list of 3D coordinates (x, y, v) along a filament skeleton. We estimated the velocity coordinates using moment-2 values. There are three user-defined parameters: 
\begin{enumerate}
    \item \texttt{npix}: the number of pixels used for perpendicular cuts
    \item \texttt{avg\_len}: the length of a segment in pc
    \item \texttt{avg\_sep}: the separation of perpendicular slices in pc
\end{enumerate}

We fixed \texttt{npix} to be 30, \texttt{avg\_len} to be $3\theta_\mathrm{B}$ and \texttt{avg\_sep} to be $0.5\theta_\mathrm{B}$, where $\theta_\mathrm{B}$ is the data cube's angular resolution.     

\begin{figure*}
    \includegraphics[width=1\textwidth]{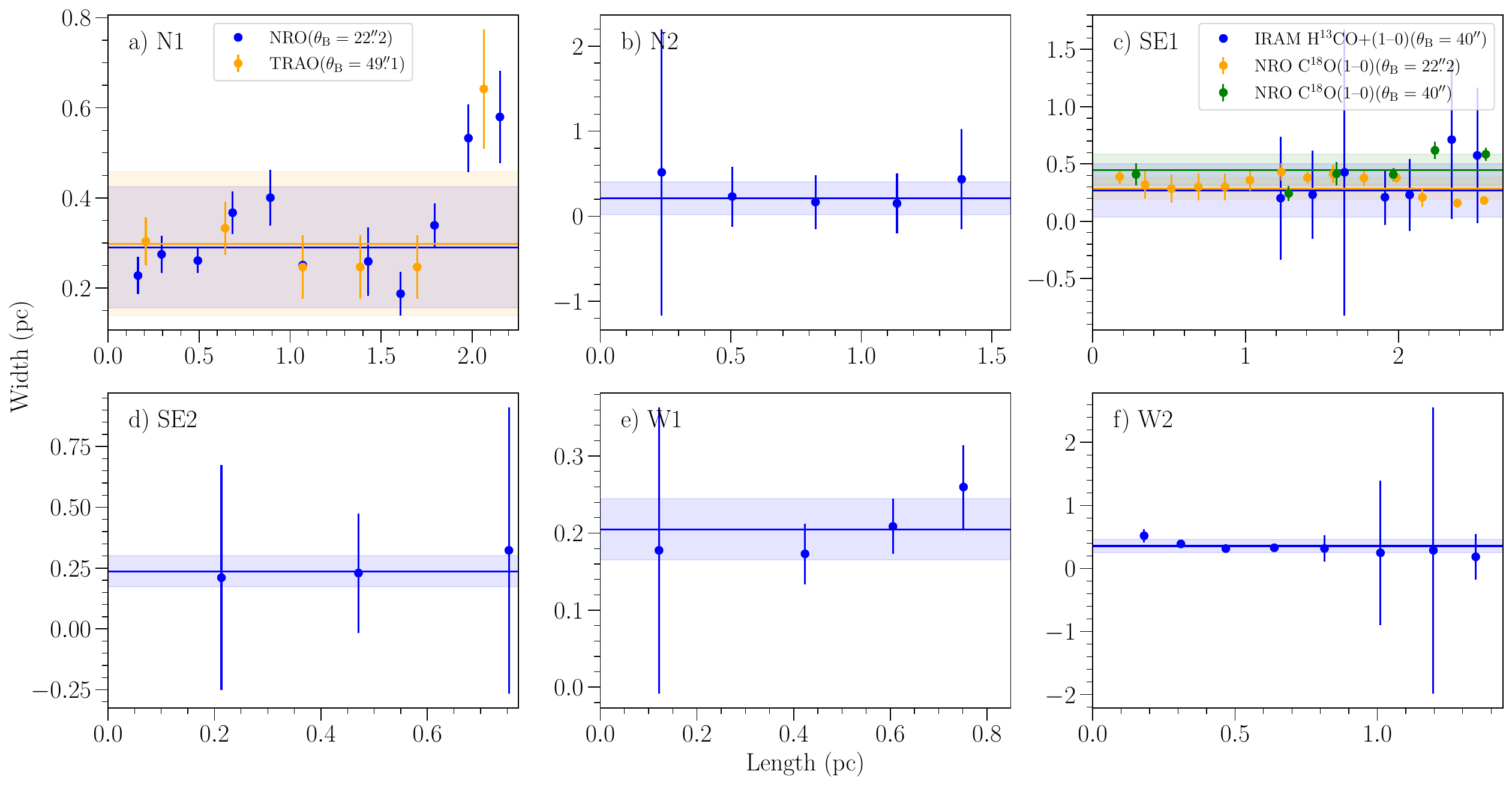}
    \caption{Filament widths derived from \texttt{FilChaP}. Panel (a) Comparison of the same tracer at different resolutions (TRAO vs. NRO). Panel (c) Comparison of different tracers (C$^{18}$O vs. H$^{13}$CO$^+$). Horizontal lines mark the weighted mean widths, and shaded rectangles indicate the 1$\sigma$ scatter. Colours match between points, lines, and shaded regions.}
    \label{fig: filchap_width_profile}
\end{figure*}

We report the representative filament width by computing the weighted mean of these individual FWHM measurements. Width uncertainties were calculated as the unweighted standard deviation of the FWHM measurements, reflecting the variability in filament width observed along its length. Additionally, to compare width estimation using different tracers and angular resolutions, we applied this procedure to N1 using TRAO C$^{18}$O data, SE1 using NRO C$^{18}$O data, and SE1 using NRO C$^{18}$O data smoothed to IRAM H$^{13}$CO$^+$ resolution. The results are presented in Fig.~\ref{fig: filchap_width_profile}.

\texttt{FilChaP} calculates filament length by summing the physical separations between consecutive pixel coordinates along the skeleton. The uncertainty in lengths is estimated by combining the uncertainties from distance and beam size in quadrature.

\subsection{Filament Profiles} 
\label{sec: fil_profile_derivation}
We first defined circular regions along the filament skeleton, each with a diameter equal to the filament width and a separation equal to the circle radius. Within each region, we fitted a single-Gaussian profile to each pixel using the \texttt{fit\_lines} procedure in the \texttt{specutils} package. This allowed us to estimate $T_\mathrm{peak}$, $\vlsr$ and $\sigma_\mathrm{obs}$. We only fitted strong spectra ($>8\sigma$). For C$^{18}$O detected filaments, we smoothed the spectra to a 0.15\kms\! velocity resolution before fitting, due to the complex profiles. For each region, we then calculated the weighted mean and standard deviation of these parameters. The weight used for the mean was a combination (in quadrature) of the fitting error and the root mean squared residual (RSS) of the fit. The mean and standard deviation of $T_\mathrm{peak}$, $\vlsr$ and $\sigma_\mathrm{obs}$ are shown in Fig.~\ref{fig: filament_T_vlsr_dv_profile} as data points and uncertainties. These values were then used to derive the following physical parameters.

\subsection{Column Density}
We first derived the column densities of the species where filaments were detected (C$^{18}$O and H$^{13}$CO$^+$) by assuming LTE and optically thin scenario, using Equations 89, 90 from \cite{Mangum_2015}.

\begin{multline}
    N_{\text{tot}}^\text{thin}=\frac{3h}{8\pi^3\mu^2J_uR_i} \left(\frac{kT_\text{ex}}{hB} + \frac{1}{3}\right)\text{exp}\left(\frac{E_u}{kT_\text{ex}}\right) \left[{\text{exp}\left(\frac{h\nu}{kT}\right)-1}\right]^{-1} \\
    \frac{\int{T_\text{B}\,dv}} {f[J_\nu(T_{\text{ex}})-J_\nu(T_{\text{bg}})]} \;\text{cm}^{-2}
\end{multline}

Where $J_\nu(T)=\frac{h\nu/k}{e^{h\nu/kT}-1}$ is the equivalent black body temperature in the Rayleigh--Jeans approximation. The line frequency ($\nu$), species dipole moment ($\mu$), and rotational constant ($B$) were obtained from CDMS: 
\begin{enumerate}
\item C$^{18}$O: $\nu=109.782182$\,GHz, $E_u=5.27$\,K, $\mu=0.11\times10^{18}$\,esu\,cm and $B=5.49\times10^{-2}$\,THz; 
\item H$^{13}$CO$^+$: $\nu=86.754288$\,GHz, $E_u=4.16$\,K, $\mu=3.9\times10^{18}$\,esu\,cm and $B=4.34\times10^{-2}$\,THz; 
\end{enumerate}
Both lines have relative intensity, $R_i$=1 because they do not have hyperfine structure. We acknowledge that $\rm C^{18}O$ may not always be optically thin in the W40 cloud \citep{tursun2024}, which could lead to an underestimation of the calculated column density in some regions.

We assumed $T_{\textrm{ex}} = T_{\textrm{dust}}$. This is justified at high densities ($n(\mathrm{H_2})\ge10^5\,\mathrm{cm^{-3}}$) where the dust and gas are thermally coupled \citep{juvela2011,goldsmith2001}. This assumption is more valid for H$^{13}$CO$^+$ filaments than C$^{18}$O filaments due to lower C$^{18}$O effective density. $T_{\textrm{dust}}$ was obtained from the \emph{Herschel} dust temperature map \citep{konyves2015}. Similar to Section~\ref{sec: fil_profile_derivation}, we computed the mean and standard deviation of $T_{\textrm{dust}}$ for each circular region. $\int{T_\text{B}\,dv}$ was calculated 
using the equation for area under a Gaussian: $\int{T_\text{B}\,dv}=T_\mathrm{peak}\sigma_\mathrm{obs}\sqrt{2\pi}$, using the $T_\mathrm{peak}$ and $\sigma_\mathrm{obs}$ values obtained in Section~\ref{sec: fil_profile_derivation}. The uncertainties in the  column densities were estimated using Monte Carlo simulations that considered the standard deviations of $T_{\textrm{dust}}$, $T_\mathrm{peak}$, $\sigma_\mathrm{obs}$. 

Finally, we converted $N_\mathrm{species}$ to $N_\mathrm{H_2}$ using a conversion factor $X_\mathrm{species}=N_\mathrm{species}/N_\mathrm{H_2}$. Since we also have the \emph{Herschel} column density map $N\mathrm{_{H_2}^{Herschel}}$, we estimated the mean $X_\mathrm{species}$ of each filament by dividing $N_\mathrm{species}$ by $N\mathrm{_{H_2}^{Herschel}}$. First, the background $N\mathrm{_{H_2}^{Herschel}}$ value (5.15$\times10^{21}$cm$^{-2}$) was subtracted. The background was estimated from the mean value in a 1\arcmin\! diameter circular region centred at ($l$=28.73\degr, $b$=3.36\degr), where no significant emission is observed. The values for $X_\mathrm{C^{18}O}$ range from $(3.11 \textrm{ to }3.94)\times10^{-7}$, consistent with \cite{shimoikura2019} for the same region, but double the value reported in other regions \citep[$1.7\times10^{-7}$ in $\rho$ Oph and Taurus, ][]{frerking1982}. The values of $X_\mathrm{H^{13}CO^+}$ range from $(1.82 \textrm{ to }3.24)\times10^{-11}$. We adopted a common mean $X_\mathrm{species}$ value for each filament. $N_\mathrm{H_2}$ was then obtained by dividing $N_\mathrm{species}$ by this adopted $X_\mathrm{species}$, and uncertainties were propagated from $N_\mathrm{species}$ uncertainties.

\subsection{Line Mass and Total Mass}
First, we estimated the surface density ($\Sigma^0_\mathrm{fil}$), representing the mass per unit area along the filament from $ N_\mathrm{H_2} $:

\begin{equation}
    \Sigma^0_\mathrm{fil} = N_\mathrm{H_2} \cdot \mu_\mathrm{H_2} \cdot m_\mathrm{H}
\end{equation}

Where $\mu_\mathrm{H_2}=2.8$ is the effective molecular weight per hydrogen molecule, and $ m_\mathrm{H} =1.0079$u is the mass of a hydrogen atom. Assuming that the filaments have a Gaussian radial column density profile, The surface density profile was then multiplied by the area under a Gaussian with an FWHM equal to the filament width. which yielded the line mass profile ($M_\mathrm{line}$) along the filament:

\begin{equation}
    M_\text{line} = \Sigma^0_\mathrm{fil} \cdot A_\text{Gaussian}
\end{equation}

where $ A_\text{Gaussian} = 1.064 \cdot \text{FWHM}$.

The uncertainties of line masses were propagated from the width and $N_\mathrm{H_2}$ uncertainties.

The total mass of a filament ($M_\mathrm{filament}$) was calculated by summing the product of line mass at each point with the separation  ($\Delta s_i$) between two consecutive points on the skeleton:

\begin{equation}
    M_\mathrm{filament}=\sum_iM_\text{line}(i)\Delta s_i
\end{equation}

\subsection{Velocity Dispersions and Virial Line Mass} \label{sec: fil_dispersions_derivation}

The observed velocity dispersion $\sigma_\mathrm{obs}$ is due to thermal and non-thermal motions. We calculated the thermal $(\sigma_\mathrm{th})$, non-thermal $(\sigma_\mathrm{nt})$ and total $(\sigma_\mathrm{total})$ velocity dispersions from $\sigma_\mathrm{obs}$ using the following equations from \cite{hacar2022}:

\begin{equation}
    \sigma_\mathrm{th}=c_s=\sqrt{\frac{k_\mathrm{B} T_\mathrm{k}}{\mu m_\mathrm{H}}}
\end{equation}

\begin{equation}
    \sigma_\mathrm{nt}=\sqrt{\sigma_\mathrm{obs}^2-\frac{k_\mathrm{B} T_\mathrm{k}}{\mu_\mathrm{species} m_\mathrm{H}}}
\end{equation}

\begin{equation}
    \sigma_\mathrm{total}=\sqrt{\sigma_\mathrm{nt}^2+\sigma_\mathrm{th}^2}
\end{equation}

Here, $\mu$=2.33 is the mean molecular weight per free particle in a molecular cloud, $\mu_\mathrm{C^{18}O}=30$ and $\mu_\mathrm{H^{13}CO^+}=30$ 
are the molecular weights for the respective species. $\sigma_\mathrm{th}$ is also equal to the isothermal sound speed, $c_s$. Since $T\mathrm{_{ex}}$ could not be estimated due to the self-absorption in $^{12}$CO and HCO$^+$, we assumed $T_k=T_\mathrm{dust}$, the dust temperature from \emph{Herschel}, assuming the gas and dust are well mixed in the filaments. We calculated the ratio $\sigma_\mathrm{nt}/c_s$ to classify non-thermal motions within the filaments as subsonic ($\sigma_\mathrm{nt}/c_s<1$), transonic ($1<\sigma_\mathrm{nt}/c_s<2$), or supersonic ($\sigma_\mathrm{nt}/c_s>1$) \citep{Chung_2019, arzoumanian2019}. 

The virial line mass ($M\mathrm{_{line}^{vir}}$) was derived following the equation by \cite{wang2014}, which considers the equilibrium between gravity, thermal, and non-thermal pressures:

\begin{equation}
    M\mathrm{_{line}^{vir}}=\frac{2\sigma^2_\text{total}}{G}=465\left(\frac{\sigma_\text{total}}{1 \text{ km s}^{-1}}\right)^2 \text{ M}_{\sun} \text{ pc}^{-1}
\end{equation}

\subsection{Velocity Gradient}
We derived the velocity gradient of each filament by performing a linear fit to the $\vlsr$ profile and calculating the slope. To evaluate the significance of this fit, we compared its Akaike Information Criterion (AIC) to that of a horizontal line fit (which assumes no velocity gradient). The AIC is a statistical measure used to evaluate the quality of different models. We only retained fits where the difference in AIC was greater than two.

\section{Physical Parameters of the clumps} 
\label{sec: clump_parameters_derived}

\subsection{Major and Minor Axes}
To estimate the major and minor axes of each clump, we projected the 3D dendrogram mask onto the position-position (PP) plane by collapsing along the velocity axis. The resulting 2D binary mask was labelled to identify contiguous regions, and the morphological properties were computed using the \texttt{regionprops} function from \texttt{scikit-image}. The lengths of the major and minor axes were taken from the ellipse that best matches the clump's shape, as defined by its pixel distribution.

\subsection{Column Density}
For each clump, we fitted all pixels with strong emission ($T_\mathrm{peak}>3\sigma$) enclosed within its boundary with a single-component hyperfine structure profile under the assumption of LTE. This assumes a common $T_\mathrm{ex}$ and line width for all hyperfine lines. The fitting was performed using the \texttt{HfS} code developed by \citet{Estalella_2017}, which simultaneously fits four independent parameters:

\begin{enumerate}
    \item $A_m^*=A(1-e^{-\tau_m})$: the observed peak intensity of the main hyperfine line, where $\tau_m$ is its optical depth;
    \item $\vlsr$: the LSR velocity of the main hyperfine line;
    \item $\Delta V$: the line width of each hyperfine line;
    \item $\tau^*_m=(1-e^{-\tau_m})$
\end{enumerate}

The spectra were spectrally smoothed to 0.15\kms\! before the fitting.

The reported means and standard deviations were calculated over all fitted spectra and are taken as the representative values and uncertainties.

$T_\mathrm{ex}$ was then calculated using the radiative transfer equation:

\begin{equation}
    A=f[J_\nu(T_\mathrm{ex})-J_\nu(T_\mathrm{bg})]
\end{equation}

where $f$ is the beam filling factor (assumed to be 1), and $J_\nu(T)$ is the Planck-corrected brightness temperature:

\begin{equation}
    J_\nu(T) = \frac{h\nu/k}{e^{h\nu/kT} - 1}.
\end{equation}

The uncertainty in $T_\mathrm{ex}$ is estimated from Monte Carlo simulations considering the uncertainties of $A_m^*$ and $\tau^*_m$.

The total column density of N$_2$H$^+$ is calculated using Equation (95) of \citet{Mangum_2015}:

\begin{equation}
    N\mathrm{_{tot}(N_2H^+)}=\frac{3h}{8\pi^3\mu^2}\frac{Q_\mathrm{rot}}{J_uR_i}\exp{\left(\frac{E_u}{kT_\mathrm{ex}}\right)}\left[\exp{\left(\frac{h\nu}{kT_\mathrm{ex}}\right)}\right]^{-1}\int{\tau_\nu\,d\nu}
\end{equation}

where $Q_\mathrm{rot}$ is the rotational partition function, $R_i\!=\!7/27$ is the relative intensity of the main hyperfine line ($F_1\!=\!2\!-\!1, F\!=\!3\!-\!2$) (Table~\ref{tab: hfs_intensity_ratio}), and $\int{\tau_\nu\, d\nu}$ represents the total opacity of the main hyperfine line. The parameters $\nu$, $\mu$ and $B$ for N$_2$H$^+$ were obtained from CDMS: $\nu=91.3173770$\,GHz, $E_u=4.38$\,K, $\mu=3.37\times10^{18}$\,esu\,cm and $B=4.66\times10^{-2}$\,THz. The uncertainty is estimated from Monte Carlo simulations considering the uncertainties of $\tau_m$ and $T_\mathrm{ex}$.

\begin{table*}
\centering
\caption{Hyperfine structure splitting transitions resolvable by the spectrometer.}
\label{tab: hfs_intensity_ratio}
\begin{tabular}{clcrr}
\hline
Molecule & 
Rotational Transition & 
Hyperfine Transition & 
Velocity Offset (\kms) & 
Relative Strength \\
\hline
C$_2$H     & N = 1 -- 0, J = 3/2 -- 1/2 & F = 1 -- 0                 & -40.13       & 5/16     \\
           &                            & F = 2 -- 1                 & 0            & 10/16    \\ 
           &                            & F = 1 -- 1                 & 112.59       & 1/16     \\
\hline
CN         & N = 1 -- 0, J = 3/2 -- 1/2 & F = 3/2 -- 3/2             & -47.38       & 3/20     \\ 
           &                            & F = 1/2 -- 1/2             & -22.91       & 3/20     \\
           &                            & F = 5/2 -- 3/2             & 0            & 10/20     \\
           &                            & F = 3/2 -- 1/2             & 7.53         & 4/20     \\
\hline
HCN        & J = 1 -- 0                 & F = 0 -- 1                 & -7.06        & 1/9       \\
           &                            & F = 2 -- 1                 & 0            & 5/9       \\
           &                            & F = 1 -- 1                 & 4.8          & 3/9       \\
\hline                                  
H$^{13}$CN & J = 1 -- 0                 & F = 0 -- 1                 & -7.25        & 1/9       \\
           &                            & F = 2 -- 1                 & 0            & 5/9       \\
           &                            & F = 1 -- 1                 & 4.97         & 3/9       \\
\hline                 
N$_2$H$^+$ & J = 1 -- 0                 & F$_1$ = 0 -- 1, F = 1 -- 2 & -8.01        & 3/27     \\
           &                            & F$_1$ = 2 -- 1, F = 1 -- 1 & -0.61        & 3/27     \\
           &                            & F$_1$ = 2 -- 1, F = 3 -- 2 & 0            & 7/27     \\
           &                            & F$_1$ = 1 -- 1, F = 2 -- 1 & 0.96         & 5/27     \\
           &                            & F$_1$ = 1 -- 1, F = 1 -- 0 & 5.55         & 3/27     \\
           &                            & F$_1$ = 1 -- 1, F = 2 -- 2 & 5.98         & 5/27     \\
           &                            & F$_1$ = 1 -- 1, F = 0 -- 1 & 6.94         & 1/27     \\
\hline
\end{tabular}
\medskip
\begin{minipage}{0.8\textwidth}
Quantum numbers of the transition are based on the notation provided in CDMS. The references of the relative strengths are: C$_2$H\citep{li2012}; CN\citep{hwang2024}; HCN\citep{Goicoechea_2022}; H$^{13}$CN\citep{ikeda2002a}; N$_2$H$^+$\citep{Mangum_2015}.
\end{minipage}
\end{table*}

\subsection{Velocity Dispersions}

The thermal $(\sigma_\mathrm{th})$, non-thermal $(\sigma_\mathrm{nt})$ and total $(\sigma_\mathrm{total})$ velocity dispersions were calculated similarity to that of filaments in Section~\ref{sec: fil_dispersions_derivation}, with $\mu_\mathrm{N_2H^+}=29$ is the molecular weight of $\mathrm{N_2H^+}$ and $\sigma_\mathrm{obs}$ obtained from the line width of the fitting results. As with the filaments, we assumed $T_k=T_\mathrm{dust}$ and used the mean \emph{Herschel} $T_\mathrm{dust}$ value within the clump boundary. Errors are propagated from the $\sigma_\mathrm{obs}$ fitting uncertainties and the standard deviation of $T_\mathrm{dust}$.

\subsection{Clump Mass}

The clump mass ($M_\mathrm{clump}$) was calculated by first converting $N_\mathrm{N_2H^+}$ to $N_\mathrm{H_2}$ assuming an abundance ratio of $X_\mathrm{N_2H^+} = N_\mathrm{N_2H^+}/N_\mathrm{H_2}=7\times10^{-10}$, the average of values found by \cite{shimoikura2019} (see their Figure~10c), and then applying the equation:

\begin{equation}
    M_\mathrm{clump}=\mu_\mathrm{H_2}m_\mathrm{H}N_\mathrm{pixel}A_\mathrm{pixel}N_\mathrm{H_2}
\end{equation}

Where $N_\mathrm{pixel}$ is the number of pixels bounded by \textsc{astrodendro} boundary, 
and $A_\mathrm{pixel}$ is the area of one pixel. Uncertainties were propagated from the column density uncertainties. We note that the mass may have additional uncertainties due to the unresolved TRAO observation and the conversion factor. A small additional uncertainty comes from slight variations in the \textsc{astrodendro} boundary introduced by parameter choices.

\subsection{Virial Parameter}

To assess the gravitational stability of the clumps, we calculated the virial parameter ($\alpha_\mathrm{vir}$) using the formulation from \citep{bertoldi1992}. This formulation assumes a spherical clump with uniform density and neglects external pressure or magnetic field support: 

\begin{equation}
    \alpha_\mathrm{vir}=\frac{5\sigma^2R}{GM}
\end{equation}

where $M$ is the clump mass, $G$ is the gravitational constant, and $\sigma$ is the total one-dimensional velocity dispersion. We adopt $\sigma=\sigma_\mathrm{total}$, which includes both thermal and non-thermal contributions in quadrature. The effective radius $R$ is defined as $R=R_\mathrm{eq}=\sqrt{A/\pi}$, following \citep{rigby2019}, where $A$ is the projected area of the clump. The area is estimated assuming an ellipsoidal shape based on the clump's major and minor axes.

According to the criterion proposed by \citet{chen2019a}, clumps with $\alpha_\mathrm{vir} \leq 2$ are considered gravitationally bound, while those with $\alpha_\mathrm{vir} > 2$ are likely unbound or pressure-supported.

\section{HCN hyperfine anomaly}
\begin{figure*} 
    \centering
    \includegraphics[width=1\textwidth]{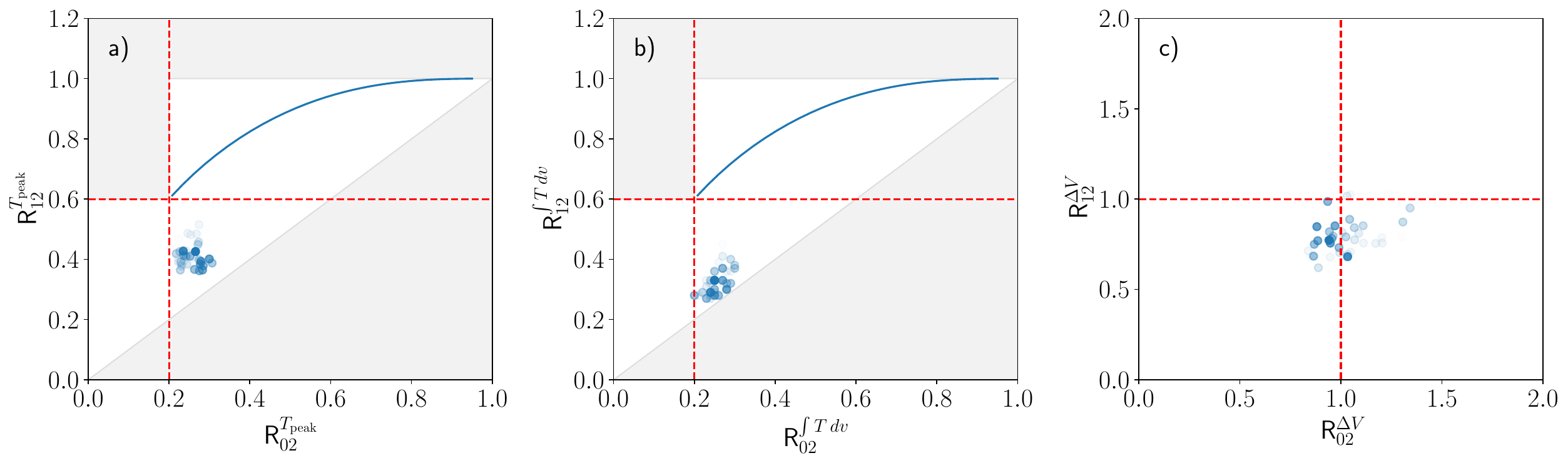}
    \caption{HCN $R_{12}$--$R_{02}$ space plotted in terms of (a) $T_\mathrm{peak}$, (b) $\int T\,dv$, and (c) $\Delta V$, following Figure~5 in \citet{Goicoechea_2022}. $R_{12}$ refers to the ratio of F(1--1) line to the F(2--1) line, while $R_{02}$ refers to the ratio of F(0--1) line to the F(2--1) line. Red dotted lines indicate the LTE ratios in the optically thin limit. Blue curves in panels (a) and (b) show the LTE ratios as optical depth $\tau$ increases. Blue points correspond to fits for the 5\kms\! component. The opacity of the scatter points is scaled by the residual sum of squares (RSS) of the fits to de-emphasize poor fits.}
    \label{fig: hcn_hfs_ratio}
\end{figure*}

To investigate deviations of the HCN hyperfine structure from LTE, we analysed the ratios of $T_\mathrm{peak}$, $\int T\,dv$, and $\Delta V$ between the satellite lines and the main line for the 5\kms\! component 
. We performed simultaneous three-component Gaussian fits on a pixel-by-pixel basis using the \texttt{fit\_lines} routine in the \texttt{specutils} package. The 10\kms\! component was excluded from the analysis because its spectra frequently exhibit a secondary component at 5\kms\! along the line of sight, which causes blending between the F(1--1) line of the 5\kms\! component and the F(2--1) line of the 10\kms\! component. The 3\kms\! component was also excluded from the analysis due to the frequent non-detection of the F(1--1) line (see Section~\ref{subsec: molecular_cloud_distribution}). The results are presented in the $R_{12}$--$R_{02}$ space following the format of Figure~5 in \citet{Goicoechea_2022}, as shown in Fig.~\ref{fig: hcn_hfs_ratio}.
 
\bsp	
\label{lastpage}
\end{document}